\documentclass[aps,pre,twocolumn,showpacs]{revtex4-1}
\usepackage[dvips,pdftex]{graphicx}
\usepackage{hyperref}
\usepackage{amsmath}
\usepackage{amssymb}
\usepackage{mathrsfs}
\usepackage[figuresright]{rotating}
\usepackage{braket}
\usepackage{bm}
\usepackage{amsmath,amssymb}
\usepackage{color}
\usepackage{url}
\usepackage{algorithm}
\usepackage[titletoc]{appendix}
\usepackage{bbm}
\usepackage{bbold}
\usepackage{mathrsfs}
\usepackage[capitalize]{cleveref}
\usepackage{bm,bbm,dsfont}
\usepackage{comment}

\usepackage[utf8]{inputenc}

\begin{document}
\title{Asymptotic densities of planar L\'{e}vy walks:
a non-isotropic case}

\author{Yu. S. Bystrik$^1$ and S. Denisov$^2$}
\affiliation{$^1$  Institute of Applied Physics, National Academy of Sciences of Ukraine, Petropavlivska Street 58, 40000 Sumy, Ukraine}
\affiliation{$^2$ Department of Computer Science, Oslo Metropolitan University, N-0130 Oslo, Norway}
\date{
\today
}

\begin{abstract}
L\'{e}vy walks are a particular type of continuous-time random walks which results in a super-diffusive spreading of an initially localized packet.
The original one-dimensional model has a simple schematization %based on two complementary events
that is based on starting a new unidirectional motion event  either in the positive or in the negative direction. We consider two-dimensional generalization of L\'{e}vy walks in the form of the so-called XY-model. It describes a particle moving with a constant velocity along one of the four basic directions and randomly switching between them when starting a new motion event. We address  the ballistic regime and derive solutions for the asymptotic density profiles. The solutions have a form of first-order integrals which can be evaluated numerically. For  specific values of  parameters we derive an exact expression.
%Finally, we evaluate different spatial asymptotics of the density profiles.
The analytic results  are in perfect agreement with the results of finite-time numerical samplings.
\end{abstract}
\maketitle

\section{Introduction}
\label{sec:1}

The idea of L\'{e}vy walks (LWs) \cite{yosi1,yosi2} can be sketched as follows: A particle moves, straightforwardly and with the constant velocity $v_0$, for some time $\tau_i$, then stops, changes,~instantaneously and randomly, the direction of its motion, and starts to move along the newly chosen direction. The particle is launched from the origin at the initial instant of time and the process is iterated  until the time reaches the set threshold $t$, $\sum_{i=1}^{N} \tau_i + \bar{\tau}_{N+1} = t$, $0 < \bar{\tau}_{N+1} < \tau_{N+1}$ (that is, the last motion event is stopped once the time threshold is reached). The duration  of a motion event is drawn from a probability density function (pdf) with a slowly decaying power-law tail, $\psi(\tau) \propto \tau^{-1-\gamma}$,  $0 < \gamma < 2$.
%The most popular  quantifier of the process (which highlights the anomalous character of the spreading) is the  mean squared displacement (MSD), $\langle r^2(t) \rangle \propto t^{\gamma}$, with  $\gamma = 3- \gamma$~ \cite{yosi1,yosi2}.
During the last two decades,  this simple -- at first glance -- model has  found applications in different fields, ranging from physics and chemistry to biology and sociology, as an instrument to describe and understand complex transport phenomena~\cite{rmp}.

Most of the existing theoretical results were derived for one dimensional LW models \cite{rmp}. Although the 1$d$ set-up allows for a lot of flexibility in tailoring of  a particular experiment-relevant model, the geometry of the resulting process is simple: the particle  moves either to the right or to the left at any instant of time. Generalization of this scheme  to 2$d$  is not straightforward and several models have been proposed ~\cite{yosi2,prl2016}, with two of them being most intuitive.

In the \textit{uniform model}~\cite{prl2016}, the direction of the next flight is determined by choosing, randomly and uniformly, a point on a unit circle (on the surface of the unit sphere $\mathcal{S}^d$ in the $d$-dimensional case~\cite{yosi2,marcin2016,marcin2017,fouxon2017}). The resulting process is spatially isotropic and this allows to reduce the set of spatial variables to a single one,  $r = \vert \textbf{r}\vert$.

\begin{figure}[t]
	\includegraphics[angle=0,width=0.49\textwidth]{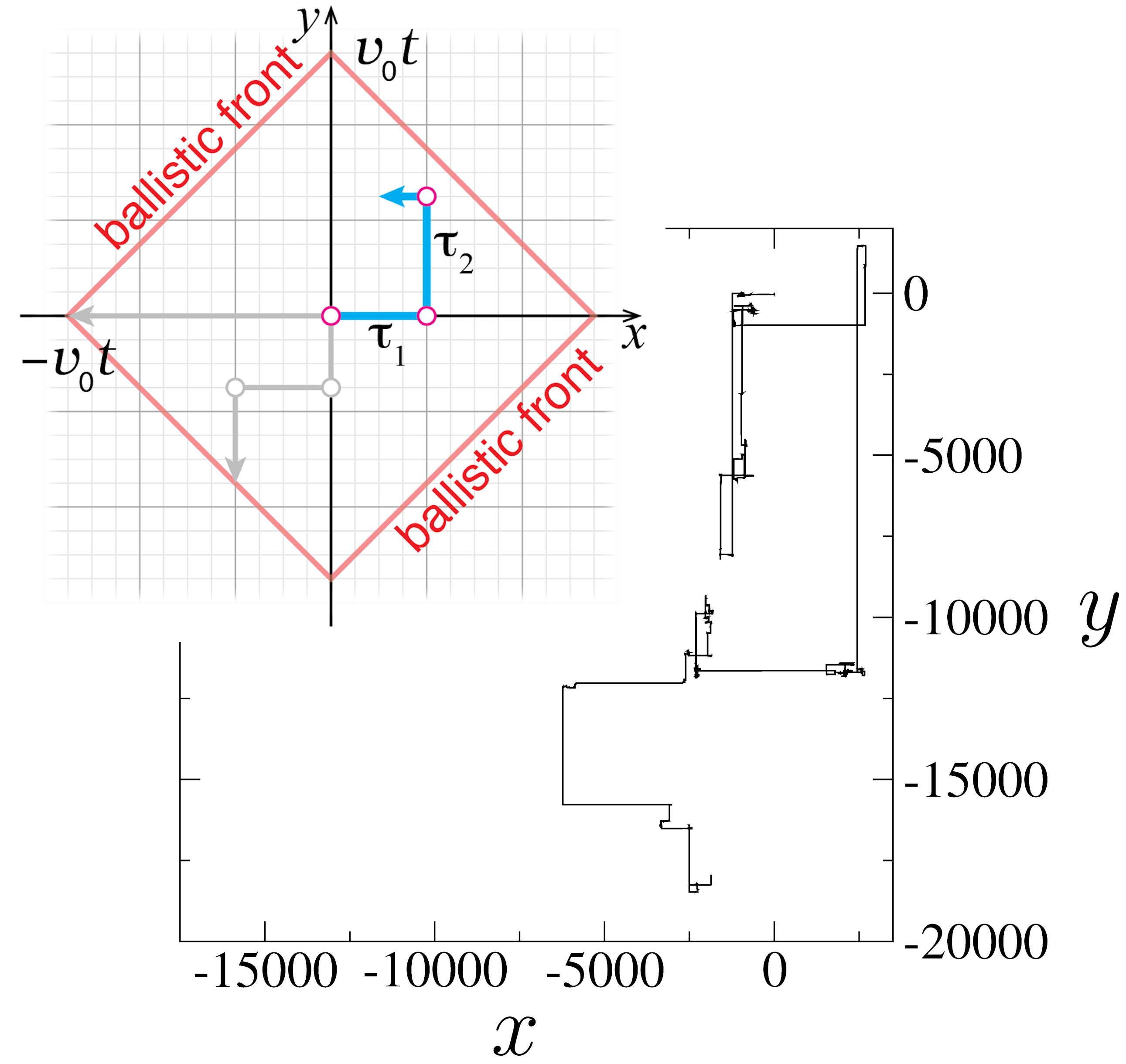}
\caption{\textbf{XY-model of planar L\'{e}vy walks.} A particle is allowed to move, with a speed $v_0$, only along one Cartesian axis at a time, which is chosen randomly at the re-orientation points $\circ$. The ballistic front is determined by the square $|x|+|y|=v_0t$. Geometry of the process imparts the shape of the corresponding trajectory which exhibits a distinctive rectangular web-like pattern with long ballistic re-locations along the two axes. The parameters here are $\gamma = 1/2$, $\upsilon_0 = 1$ and $\tau_0=1$.}
\label{fig:1}
\end{figure}

In the \textit{XY-model}~\cite{yosi3,prl2016}, the motion of the particle is restricted to four basic Cartesian directions; see Fig.~\ref{fig:1}. When initiating  a new motion event, one has to roll a four-sided die \cite{dice}, draw duration $\tau_i$, and then set the particle into a ballistic motion along the corresponding direction. The resulting process is essentially non-isotropic and that is %feature
imprinted in the shape
of  pdf $P(\textbf{r},t)$ specifying the probability of finding the particle at a  vicinity of point $\textbf{r}$ at time $t$~\cite{prl2016,fouxon2017}. The XY-model is not just an abstract mathematical construction. For example, it reproduces Hamiltonian kinetics in egg-crate potentials~\cite{yosi3} and in infinite horizon billiards~\cite{zarfaty1}. Depending on the symmetry of a potential or size of the scatterers in a billiard, the motion can be restricted to four, eight, or larger even number of  basic directions ~\cite{cristadoro}. The XY-model can be generalized to reproduce kinetics of these  systems~\cite{zarfaty2}.

In the ballistic regime, $0 < \gamma < 1$,  mean flight time $\langle \tau \rangle = \int_{0}^{\infty}\tau \psi(\tau)  \,d\tau$ diverges and the  mean squared displacement (MSD) of the corresponding LW processes exhibit universal ballistic scaling, $\langle r^2(t) \rangle = \int_{0}^{t} r^2 P(\textbf{r},t) \,d\tau\propto t^{2}$.  A method to compute
asymptotic pdf's
%$P(x,t)$
for one-dimensional ballistic L\'{e}vy walks was presented in  Ref.~\cite{bal1}. Consequently, asymptotic pdf's of the uniform model
%, both in 2- and 3-$d$,
were evaluated in Ref.~\cite{marcin2016}.

Here we advance further along this line and address  ballistic regime of the XY-model. Evidently, the corresponding spatially non-isotropic  spreading is more complex than the  one obtained with the uniform model. Remarkably, as we demonstrate, even in this case it is possible to compute the asymptotic densities  and derive 
%for them explicit 
analytical expressions.
%To obtain these results, we use several techniques which are not conventional in the field of LW studies.
%for $\gamma = \frac{1}{2}$.
%We illustrate these analytical findings with numerical simulations.

%\textit{The paper is organized as follows. 
%In Sec.~\ref{sec:2}, we describe the LWs model, basic equations and formulate
%the problem of the asymptotic probability density function (pdf). 
%In Sec.~\ref{sec:3}, we find pdf  in terms of  first-order integrals. Its alternative %representation, containing integrals with  $\gamma$-stable L\'{e}vy distribution,  are %obtained in Sec.~\ref{sec:4}.
%In Sec.~\ref{sec:5}, we develop a method for the numerical evaluation of the re-scaled %pdf and the averaged (over the bin) pdf, and compare theoretical results with the %corresponding sampling.
%In Sec.~\ref{sec:6}, an exact expression of pdf for a specific value
%of the tail parameter is derived.
%Our main findings are summarized in Sec.~\ref{sec:7}. Finally, auxiliary
%derivations are presented in the Appendix \ref{sec:AppA} and \ref{sec:AppB}.}

\section{Model and basic equations}
\label{sec:2}

Following the basic idea of LWs \cite{yosi2,rmp}, we consider a particle which moves with constant velocity  $v_0$ and performs instantaneous re-orientations at random instants of time. The time between two consequent re-orientation events is a random variable distributed according to pdf
\begin{equation}
    \psi(\tau)=\frac{1}{\tau_0}\frac{\gamma}{(1+\tau/\tau_0)^{1+\gamma}},\quad 0<\gamma<1,
    \label{fPDF}
\end{equation}
where $\tau_0>0$. The re-orientation process is determined by pdf $h(\mathbf{v})$ which specifies the direction of vector $\mathbf{v}$, $\vert \mathbf{v} \vert = v_0$.

The particle starts from the origin at the initial instant of time. The probability
to have the particle moving  without re-orientation up to time $t$ is  $\Psi(t)=\int_{t}^{\infty}\mathrm{d}\tau \psi(\tau)$. Pdf $P(\mathbf{r}, t)$, after being transformed into the Fourier-Laplace domain, obeys the equation
\begin{equation}
    P(\mathbf{k},s)=\frac{\int \mathrm{d}\mathbf{v}\, \Psi(s+i \mathbf{k} \cdot \mathbf{v})h(\mathbf{v})}
    {1-\int \mathrm{d}\mathbf{v}\, \psi(s+i \mathbf{k} \cdot \mathbf{v})h(\mathbf{v})},
    \label{trans_eq}
\end{equation}
where $\mathbf{k}=\{k_x, k_y\}$ and $s$ are coordinates in the two-dimensional Fourier and one-dimensional Laplace spaces,
respectively.
%to  $\mathbf{r}=\{x, y\}$ и $t$ in the original space-time domain.
%The Laplace transform is defined in the standard way, %$f(s)=\mathcal{L}\{f(t)\}=\int_{0}^{\infty} \mathrm{d}t %e^{-s t}f(t)$, while the Fourier transform for $x$ is %given by
%$f(k_{x})=\mathcal{F}_{x}\{f(x)\}=\int_{-\infty}^{\infty} \mathrm{d}x e^{-i k_{x}x} f(x)$ (and similarly for %$y$).

In the case of the XY-model, we have re-orientation pdf  $h(\mathbf{v})=[\delta(|v_x|-v_0)\delta(v_y)+\delta(v_x)\delta(|v_y|-v_0)]/4$.  The ballistic front has the form of a square defined by the equation $|x| + |y| = v_0 t$; see Fig.~\ref{fig:1}. In this case equation (\ref{trans_eq}) can be rewritten as
\begin{equation}
    P(\mathbf{k},s)=\frac{\sum_{\kappa \in K}\Psi(s+i \kappa v_0)}
    {\sum_{\kappa \in K}[1-\psi(s+i \kappa v_0)]},
    \label{trans_eq_XY}
\end{equation}
\\
where $K=\{\pm k_x, \pm k_y\}$. The structure of the equation  highlights the fact that $P(\mathbf{r},t)$ is an even (symmetric) function with respect to the space coordinates and is also invariant under permutation  $x\leftrightarrow y$. Henceforth we assume that $v_0=1$ and consistently re-normalized time which now is measure in units of space. It would be enough to replace  $t \mapsto v_0 t$ in the final expressions in order to obtain the answer for arbitrary $v_0$.

In the long-time limit, the waiting-time distribution~(\ref{fPDF}), can be approximated in the Laplace domain as
\begin{equation}
    \psi(s) \simeq 1 - \tau_0^{\gamma}\Gamma(1-\gamma) s^{\gamma}+o(s^{\gamma}).
    \label{psi_s}
\end{equation}
In the limit  $\mathbf{k}, s \rightarrow 0$ (which corresponds to both $\mathbf{r}$ and $t$ are going to infinity), we obtain from Eqs.~(\ref{trans_eq_XY}-\ref{psi_s}) the following expression:
\begin{equation}
    P_{XY}(\mathbf{k},s)=\frac{\sum_{\kappa \in K}(s+i \kappa)^{\gamma-1}}
    {\sum_{\kappa \in K}(s+i \kappa)^{\gamma}}.
    \label{eq_XY}
\end{equation}

It is noteworthy that, by using the notion of fractional material derivatives~ \cite{MBSB2002(r), SM2003, MetzlerKlafter2004}, a deterministic equation governing the evolution of the pdf in the original space, can be derived.
%\begin{equation}
%    \mathcal{F}_{z}\mathcal{L}\left\{ %\left(\frac{\partial}{\partial t}+ \frac{\partial}{\partial %z}\right)^{\gamma}\!f(z,t)\right\}=(s+ik_{z})^{\gamma}f(k_{z},%t),
%    \label{eq_frac_der}
%\end{equation}
%we can obtain from Eq.~(\ref{eq_XY}) the following fractional differential equation
%\begin{equation}
%   \sum_{z =\{\pm x, \pm y\}}\left(\frac{\partial}{\partial %t}+\frac{\partial}{\partial z}\right)^{\!\gamma}P_{XY}(x,y,t)=
%   [\delta(|x|-t)\delta(y)+\delta(x)\delta(|y|-t)]\frac{t^{-\a%lpha}}{\Gamma(1-\gamma)}.
%    \label{eq_diff}
%\end{equation}

\section{Derivation of asymptotic pdf $\mathcal{P}(\overline{x},\overline{y})$}
\label{sec:3}

We start with recasting Eq.~(\ref{eq_XY}) into
\begin{equation}
    P_{XY}(\mathbf{k},s)=Q(k_x,k_y,s)+Q(k_y,k_x,s),
    \label{P_ks}
\end{equation}
where
\begin{equation}
     Q(k_x,k_y,s)=
     \frac{(s+i k_x)^{\gamma-1}+(s-i k_x)^{\gamma-1}}
     {\sum_{\kappa \in K}(s+i \kappa)^{\gamma}}.
     \label{F_ks}
\end{equation}
It is enough therefore to find the inverse of function
 $Q(k_x,k_y,s)$ [the inverse of $Q(k_y,k_x,s)$ could be obtained by permuting $x\leftrightarrow y$].

We introduce the following two functions:
\begin{widetext}
\begin{gather}
    g_1\!\left(\frac{ik_{x}}{s},u\right) =\left[\left(1+\frac{i k_{x}}{s}\right)^{\gamma-1}+\left(1-\frac{i k_{x}}{s}\right)^{\gamma-1}\right]
    \exp\!\left\{-u\! \left[\left(1+\frac{i k_{x}}{s}\right)^{\gamma}+\left(1-\frac{i k_{x}}{s}\right)^{\gamma}\right]\right\},
    \label{g_1}
    \\
    g_2\!\left(\frac{ik_{y}}{s},u\right) =\exp\!\left\{-u\! \left[\left(1+\frac{i k_{y}}{s}\right)^{\gamma}+\left(1-\frac{i k_{y}}{s}\right)^{\gamma}\right]\right\}.
    \label{g_2}
\end{gather}
\end{widetext}

By implementing identity
%$1/\varrho=\int_{0}^{\infty}\mathrm{d}u e^{-u \varrho}~(\mathrm{Re}\, \varrho>0)$
\begin{equation}
    1/\varrho=\int_{0}^{\infty}\mathrm{d}u\, e^{-u \varrho} \,\, (\mathrm{Re}\, \varrho>0),
    \label{1/varrho}
\end{equation}
we can recast Eq.~(\ref{F_ks}) as
\begin{equation}
    Q(k_x,k_y,s)=\frac{1}{s}\int_{0}^{\infty}\mathrm{d}u~ g_1\!\left(\frac{ik_{x}}{s},u\right)
    g_2\!\left(\frac{ik_{y}}{s},u\right).~~
    \label{F_ks1}
\end{equation}
%This  integral form is more convenient for further %analysis  than Eq.~\eqref{F_ks} because it allows to %separate  variables in the Laplace domain.
By using properties of the Laplace transform for a derivative and a convolution (which we denote with $\circ$),
%$\mathcal{L}^{-1}\{s f(s)\}= \frac{\partial}{\partial %t}f(t)+f(0^+)$ и свойство свёртки Лапласа
%
%\begin{equation}
%    f_1(t)\circ f_2(t)=\int_{0}^{t}\mathrm{d}\tau %f_1(t-\tau) f_2(\tau)=
%    \mathcal{L}^{-1}\! \left\{ f_1(s)f_2(s) \right\},
%    \label{Lap_convol}
%\end{equation}
from Eq.~(\ref{F_ks1}) we obtain
\begin{eqnarray}
    Q(x,y,t)=\frac{\partial}{\partial t}\int_{0}^{\infty}\mathrm{d}u\,
    G_1(x,t,u)\circ G_2(y,t,u),
    \label{F_t(x,y)1}
\end{eqnarray}
where
\begin{align}
    G_1(x,t,u)&=\mathcal{F}^{-1}_{x}\mathcal{L}^{-1}\! \left\{
    \frac{1}{s}g_1\!\left(\frac{ik_{x}}{s},u\right)
    \right\}
    \label{G_1}, \\
    G_2(y,t,u)&=\mathcal{F}^{-1}_{y}\mathcal{L}^{-1}\! \left\{
    \frac{1}{s}g_2\!\left(\frac{ik_{y}}{s},u\right)
    \right\}.
    \label{G_2}
\end{align}
Thus we obtained the expression for  $Q(x,y,t)$ which demands not a three-step inverse transform,  $\mathcal{F}^{-1}_{x}\mathcal{F}^{-1}_{y}\mathcal{L}^{-1}$,
%(от $Q(k_x,k_y,s)$),
but a pair of two-step inverse transforms,  $\mathcal{F}^{-1}_{x}\mathcal{L}^{-1}$ and $\mathcal{F}^{-1}_{y}\mathcal{L}^{-1}$, of functions $\frac{1}{s}g_1\!\left(\frac{ik_{x}}{s}\right)$ and $\frac{1}{s}g_2\!\left(\frac{ik_{y}}{s}\right)$, respectively. To find the inverses, we
follow a procedure similar to that given in Ref.~\cite{GL2001} (see Appendix \ref{sec:AppA}) and obtain
\begin{align}
     G_1(x,t,u) &=  -\frac{1}{2\pi i\, x}  \!\lim\limits_{\,\epsilon \rightarrow 0^+}\!\!\left[g_1\!\left(-\frac{1}{x/t+ i\epsilon}, u\right)\right.
     \nonumber\\
     &\mathrel{\phantom{=}}
     \left.-g_1^*\!\left(-\frac{1}{x/t+ i\epsilon}, u\right)
     \right]
     \label{G_11}
\end{align}
and
\begin{align}
     G_2(y,t,u) &=
      -\frac{1}{2\pi i\, y}  \!\lim\limits_{\,\epsilon \rightarrow 0^+}\!\!\left[g_2\!\left(-\frac{1}{y/t+ i\epsilon}, u\right)\right.
     \nonumber\\
     &\mathrel{\phantom{=}}
     \left.-g_2^*\!\left(-\frac{1}{y/t+ i\epsilon}, u\right)
     \right].
     \label{G_21}
\end{align}

For the principal values of functions  $(1\pm\zeta)^{\gamma}$ the following holds
\begin{align}
     &\lim\limits_{\,\epsilon \rightarrow 0^+}(1+\zeta)^{\gamma}\big|_{\zeta=-1/(\xi\pm i\epsilon)}=\left|1-1/\xi\right|^{\gamma}e^{\pm i\pi \gamma \mathbbm{1}_{(0,1)}(\xi)},
    \nonumber \\
    &\lim\limits_{\,\epsilon \rightarrow 0^+}(1-\zeta)^{\gamma}\big|_{\zeta=-1/(\xi\pm i\epsilon)}=\left|1+1/\xi\right|^{\gamma}e^{\pm i\pi \gamma \mathbbm{1}_{(-1,0)}(\xi)},
    \nonumber
\end{align}
where we use the indicator function
\begin{equation}
\mathbbm{1}_{\mathcal{A}}(\xi) = \left\{\!\! \begin{array}{cl}
1,
& \xi \in \mathcal{A},
\\ [4pt]
0,
& \xi \not\in \mathcal{A}.
\end{array}
\right.
\label{a_finite}
\end{equation}

Taking into account that both functions,
$G_1(x,t,u)$ and $G_2(y,t,u)$, are even (symmetric) with respect to $x$ and $y$ [this trivially follows from the fact that functions \eqref{g_1} and \eqref{g_2} are even with respect to $k_x$ and $k_y$], we can re-write Eqs.~(\ref{G_1}) and (\ref{G_2}) in the following form
\begin{align}
     G_1(x,t,u) &= -\frac{\mathbbm{1}_{(0,t)}(|x|)}{2\pi i\, |x|}\! \left[h_{\gamma-1}\!\left(\frac{|x|}{t}\right)
     e^{-u h_{\gamma}\left(\frac{|x|}{t}\right)}\right.
     \nonumber\\
     &\mathrel{\phantom{=}}
     \left.-h_{\gamma-1}^*\! \left(\frac{|x|}{t}\right)
     e^{-u h_{\gamma}^*\left(\frac{|x|}{t}\right)}\right],
     \label{G_12}
\end{align}
\begin{align}
     G_2(y,t,u)&= -\frac{\mathbbm{1}_{(0,t)}(|y|)}{2\pi i\, |y|}\!
     \nonumber\\
     &\mathrel{\phantom{=}}
     \times\left[e^{-u h_{\gamma}\left(\frac{|y|}{t}\right)}
     -e^{-u h_{\gamma}^*\left(\frac{|y|}{t}\right)}\right],
     \label{G_22}
\end{align}
where
\begin{equation}
     h_{\gamma}(\xi)= \left|1-1/\xi\right|^{\gamma}e^{i\pi \gamma}+\left|1+1/\xi\right|^{\gamma}.
     \label{h}
\end{equation}

Substituting expressions \eqref{G_12} and \eqref{G_22} into Eq.~\eqref{F_t(x,y)1}, after some derivation, we obtain
\begin{align}
    Q(x,y,t) &=
    \frac{\mathbbm{1}_{(0,t)}(|x|+|y|)}{2\pi^2 |x| |y|} \text{Re}\,\frac{\partial}{\partial t}
     \!\int_{|y|}^{t-|x|}\!\!\mathrm{d}\tau
     h_{\gamma-1}\!\left(\frac{|x|}{t-\tau}\right)
   \nonumber\\
  &\mathrel{\phantom{=}}
  \times
  \!\left[\frac{1}{h_{\gamma}\!\left(\frac{|x|}{t-\tau}\!\right)+h_{\gamma}^*\!\left(\frac{|y|}{\tau}\right)}
  \right.
  \nonumber\\
  &\mathrel{\phantom{=}}
  \left. -\frac{1}{h_{\gamma}\!\left(\frac{|x|}{t-\tau}\right)+h_{\gamma}\!\left(\frac{|y|}{\tau}\right)}\right].
    \label{FF}
\end{align}
Finally, by substituting $\tau=(t-|x|-|y|)\eta+|y|$ and introducing notations
\begin{eqnarray}
    x_t=\frac{2|x|}{t-|x|-|y|},
    \quad
    y_t=\frac{2|y|}{t-|x|-|y|},
    \label{xt_yt}
\end{eqnarray}
pdf $P_{XY}(x,y,t)$ can be represented as
\begin{equation}
    P_{XY}(x,y,t)= Q(x,y,t)+Q(y,x,t),
    \label{P_XY}
\end{equation}
where
\begin{equation}
    Q(x,y,t)=\frac{1}{2\pi^2 |y|}
    \text{Re}\frac{\partial}{\partial t}R(x_t,y_t)
    \label{Qx}
\end{equation}
and
\begin{widetext}
\begin{align}
   R(x_t,y_t) &=  x_t^{-\gamma} \int_{0}^{1}\!\mathrm{d}\eta
    \left[(1-\eta)^{\gamma-1}e^{i\pi(\gamma-1)}+\left(1-\eta+x_t\right)^{\gamma-1}\right]~~~~~~~~~~~~~~~~~~~~~~~~~~~~~~~~~~~~~~
   \nonumber\\
    &\mathrel{\phantom{=}}
    \times\left\{
    \frac{1}{x_t^{-\gamma}\left[(1-\eta)^{\gamma}e^{i\pi\gamma}+\left(1-\eta+x_t\right)^{\gamma}\right] +y_t^{-\gamma}\left[\eta^{\gamma}e^{-i\pi\gamma}+\left(\eta+y_t\right)^{\gamma}\right]}
    \right.
   \nonumber\\
    &\mathrel{\phantom{=}}
    -\left.
     \frac{1}{x_t^{-\gamma}\left[(1-\eta)^{\gamma}e^{i\pi\gamma}+\left(1-\eta+x_t\right)^{\gamma}\right] +y_t^{-\gamma}\left[\eta^{\gamma}e^{i\pi\gamma}+\left(\eta+y_t\right)^{\gamma}\right]}
    \right\}
    \label{Rx}
\end{align}
\end{widetext}
if $|x|+|y|<t$ and $P_{XY}(x,y,t)=0$ if otherwise [henceforth we assume that $Q(x,y,t)$  and all related functions are multiplied with the indicator function, Eq.~\eqref{a_finite}]. Again, expressions for $Q(y,x,t)$ and $R(y_t,x_t)$ can be obtained from Eqs.~\eqref{Qx} and \eqref{Rx} by permuting $x\leftrightarrow y$.

\begin{comment}
Для наглядности и удобства дальнейших расчётов выпишим их
\begin{equation}
    Q(y,x,t)=\frac{1}{2\pi^2 |x|}
    \text{Re}\frac{\partial}{\partial t}R(y_t,x_t)
    \label{Qy}
\end{equation}
и
\begin{eqnarray}
  R(y_t,x_t) =
%  \!\!&=&\!\!
  y_t^{-\gamma} \int_{0}^{1}\!\mathrm{d}\eta
   \left[\eta^{\gamma-1}e^{i\pi(\gamma-1)}+\left(\eta+y_t\right)^{\gamma-1}\right]
%   \nonumber\\[4pt]
%   &&\!\!
   \times\!\left\{
   \frac{1}{x_t^{-\gamma}\left[(1-\eta)^{\gamma}e^{-i\pi\gamma}+\left(1-\eta+x_t\right)^{\gamma}\right] +y_t^{-\gamma}\left[\eta^{\gamma}e^{i\pi\gamma}+\left(\eta+y_t\right)^{\gamma}\right]}
   \right.
   \nonumber\\[4pt]
   &&\!\!-\left.
   \frac{1}{x_t^{-\gamma}\left[(1-\eta)^{\gamma}e^{i\pi\gamma}+\left(1-\eta+x_t\right)^{\gamma}\right] +y_t^{-\gamma}\left[\eta^{\gamma}e^{i\pi\gamma}+\left(\eta+y_t\right)^{\gamma}\right]}
   \right\}.
    \label{Ry}
\end{eqnarray}
При записи $R(y_t,x_t)$ для  однообразности  мы вдобавок  сделали замену $\eta \mapsto1-\eta$.
\end{comment}

It will be easier to compute $P_{XY}(x,y,t)$
if we take derivative with respect to time in Eq.~\eqref{Qx} and in  the corresponding expression for $Q(y,x,t)$.
 %\eqref{Qy},
As the result we obtain
\begin{align}
    P_{XY}(x,y,t) &= Q_1(x,y,t)+Q_2(x,y,t)
     \nonumber\\
    &\mathrel{\phantom{=}}
    +Q_1(y,x,t)+Q_2(y,x,t),
    \label{P_XY_Q12}
\end{align}
where
\begin{widetext}
\begin{align}
    Q_1(x,y,t) &=
    \frac{(1-\gamma)x_t^{1-\gamma} y_t}{4 \pi^2 |y|^2}  \int_{0}^{1}\!\mathrm{d}\eta\!\left(1-\eta+x_t\right)^{\gamma-2}
    \!\text{Re}\!\left\{
    \frac{1}{x_t^{-\gamma}\left[(1-\eta)^{\gamma}e^{i\pi\gamma}+\left(1-\eta+x_t\right)^{\gamma}\right] +y_t^{-\gamma}\left[\eta^{\gamma}e^{-i\pi\gamma}+\left(\eta+y_t\right)^{\gamma}\right]}
    \right.
    \nonumber\\
     &\mathrel{\phantom{=}}
     -\left.
     \frac{1}{x_t^{-\gamma}\left[(1-\eta)^{\gamma}e^{i\pi\gamma}+\left(1-\eta+x_t\right)^{\gamma}\right] +y_t^{-\gamma}\left[\eta^{\gamma}e^{i\pi\gamma}+\left(\eta+y_t\right)^{\gamma}\right]}
    \right\},
    \label{Q1x}
\end{align}
\begin{align}
    Q_2(x,y,t) &=
    \frac{\gamma x_t^{-\gamma} y_t}{4 \pi^2 |y|^2} \int_{0}^{1}\!\mathrm{d}\eta\!
    \left[x_t^{1-\gamma}\left(1-\eta+x_t\right)^{\gamma-1} +y_t^{1-\gamma}\left(\eta+y_t\right)^{\gamma-1}\right]
    \text{Re}\!
    \left[(1-\eta)^{\gamma-1}e^{i\pi(\gamma-1)}+\left(1-\eta+x_t\right)^{\gamma-1}\right]
    \nonumber\\
    &\mathrel{\phantom{=}}
    \times\!\left\{
    \frac{1}{\left\{x_t^{-\gamma}\left[(1-\eta)^{\gamma}e^{i\pi\gamma}+\left(1-\eta+x_t\right)^{\gamma}\right] +y_t^{-\gamma}\left[\eta^{\gamma}e^{-i\pi\gamma}+\left(\eta+y_t\right)^{\gamma}\right]\right\}^2}
    \right.
    \nonumber\\
    &\mathrel{\phantom{=}}
    -\left.
     \frac{1}{\left\{x_t^{-\gamma}\left[(1-\eta)^{\gamma}e^{i\pi\gamma}+\left(1-\eta+x_t\right)^{\gamma}\right] +y_t^{-\gamma}\left[\eta^{\gamma}e^{i\pi\gamma}+\left(\eta+y_t\right)^{\gamma}\right]\right\}^2}
    \right\}.
    \label{Q2x}
\end{align}
\end{widetext}

By introducing coordinates
\begin{equation}
    \overline{x}=\frac{1}{t}\int_{0}^{t}v(t')\mathrm{d}t'=\frac{x}{t},
    \quad
    \overline{y}=\frac{1}{t}\int_{0}^{t}v(t')\mathrm{d}t'=\frac{y}{t},
    \label{x_y_overline}
\end{equation}
for which the pdf has the form
\begin{equation}
    \mathcal{P}(\overline{x},\overline{y})=t^2P_{XY}(t\overline{x},t\overline{y},t),
    \label{P_scale}
\end{equation}
we can obtain an expression that does not depend on $t$ in the explicit way. We will not write it here; it can be obtained straightforwardly from Eq.~\eqref{P_XY_Q12} by replacing  $x \rightarrow \overline{x}$, $y \rightarrow \overline{y}$ and $t \rightarrow 1$.

\section{Alternative representation of $\mathcal{P}(\overline{x},\overline{y})$}
\label{sec:4}

Here we derive an alternative expression for  $P_{XY}(x,y,t)$ which will be used  to derive exact analytical results for $\gamma = \frac{1}{2}$ in Section~\ref{sec:6}.

First, we recast Eq.~(\ref{eq_XY}) as
\begin{align}
    P_{XY}(\mathbf{k},s) &= H(k_x,k_y,s)+H(-k_x,k_y,s)
    \nonumber\\
    &\mathrel{\phantom{=}}
    +H(k_y,k_x,s)+H(-k_y,k_x,s),
    \label{P_ks1}
\end{align}
where
\begin{equation}
    H(k_x,k_y,s)=
    \frac{(s+i k_x)^{\gamma-1}}{\sum_{\kappa \in K}(s+i \kappa)^{\gamma}}.
    \label{H_ks}
\end{equation}

We use Eq.~\eqref{1/varrho}, together with the definition of the Laplace transform of a convolution, to obtain
\begin{equation}
    H(x,y,t)=\int_{0}^{\infty}\mathrm{d}u\, H_1(x,t,u) \circ H_2(y,t,u)
    \label{H}
\end{equation}
where
\begin{align}
    H_1(x,t,u) &=
    \mathcal{F}^{-1}_{x}\mathcal{L}^{-1}\! \Big\{(s+i k_x)^{\gamma-1}
    \nonumber\\
    &\mathrel{\phantom{=}}
    \times e^{-u (s+i k_x)^{\gamma}}e^{-u (s-i k_x)^{\gamma}}\Big\},
    \label{H_1}
\end{align}
\begin{equation}
    H_2(y,t,u)=\mathcal{F}^{-1}_{y}\mathcal{L}^{-1}\! \left\{e^{-u (s+i k_y)^{\gamma}}e^{-u (s-i k_y)^{\gamma}}\right\}.
    \label{H_2}
\end{equation}
Next we use the property of the Fourier transform of a convolution (which we denote with $\bullet$) to rewrite functions (\ref{H_1}) and (\ref{H_2}) as
\begin{align}
    H_1(x,t,u) &= \mathcal{F}^{-1}_{x}\mathcal{L}^{-1}\! \left\{(s+i k_x)^{\gamma-1}e^{-u (s+i k_x)^{\gamma}}\right\}
    \nonumber\\
    &\mathrel{\phantom{=}}
    \circ \bullet_x \,
    \mathcal{F}^{-1}_{x} \mathcal{L}^{-1} \! \left\{e^{-u (s-i k_x)^{\gamma}}\right\},
    \label{H_11}
\end{align}
\begin{align}
    H_2(y,t,u) &= \mathcal{F}^{-1}_{y}\mathcal{L}^{-1}\! \left\{e^{-u (s+i k_y)^{\gamma}} \right\}
    \nonumber\\
    &\mathrel{\phantom{=}}
    \circ \bullet_y \,
    \mathcal{F}^{-1}_{y} \mathcal{L}^{-1}\! \left\{ e^{-u (s-i k_y)^{\gamma}}\right\}.
    \label{H_21}
\end{align}

It is now clear that we are dealing with one-sided $\gamma$-stable L\'evy distribution $\ell_{\gamma}(t)=\mathcal{L}^{-1}\{e^{-s^{\gamma}}\}$~\cite{SamorodnitskyTaqqu}. It is easy to see that for  $u>0$ we have
\begin{gather}
  \mathcal{L}^{-1}\! \left\{ e^{-u s^{\gamma}}\right\}= u^{-1/\gamma} \ell_{\gamma}(u^{-1/\gamma}t),
    \label{eq_levy1}
    \nonumber\\
   \mathcal{L}^{-1}\! \left\{ s^{\gamma-1}e^{-u s^{\gamma}}\right\}=\frac{t}{\gamma u}  u^{-1/\gamma} \, \ell_{\gamma}(u^{-1/\gamma} t).
    \label{eq_levy2}
    \nonumber
\end{gather}

Using  these expressions  together with
the property of a shifted inverse Laplace transform,
$\mathcal{L}^{-1}\{f(s+b)\}=e^{-bt}f(t)$, and the fact that  $\mathcal{F}_{x}^{-1}\{e^{-ik_{x}b}\}=\delta(x+b)$ (the same stands for $y$),  from Eqs.~\eqref{H_11} and \eqref{H_21} we obtain
\begin{align}
    H_1(x,t,u) &= \mathbbm{1}_{(0,t)}(|x|)\frac{u^{-2/\gamma-1}}{2\gamma}
    \nonumber\\
    &\mathrel{\phantom{=}}
    \times
    \frac{t+x}{2}
    \ell_{\gamma}\!\left(\frac{t+x}{2u^{1/\gamma}}\right)\ell_{\gamma}\!\left(\frac{t-x}{2u^{1/\gamma}}\right)
    \label{H_1_solve}
\end{align}
and
\begin{align}
    H_2(y,t,u) &=\mathbbm{1}_{(0,t)}(|y|)\frac{u^{-2/\gamma}}{2}
    \nonumber\\
    &\mathrel{\phantom{=}}
    \times
    \ell_{\gamma}\!\left(\frac{t+y}{2u^{1/\gamma}}\right)
    \ell_{\gamma}\!\left(\frac{t-y}{2u^{1/\gamma}}\right).
    \label{H_2_solve}
\end{align}

Substituting (\ref{H_1_solve}) and (\ref{H_2_solve}) into Eq.~(\ref{H}), we get
\begin{align}
   H(x,y,t) &= \frac{\mathbbm{1}_{(0,t)}(|x|+|y|)}{8\gamma}
  \int_{0}^{\infty}\mathrm{d}u\,u^{-4/\gamma-1}
   \nonumber\\
   &\mathrel{\phantom{=}}
   \times \int_{|y|}^{t-|x|}\mathrm{d}\tau\, (t-\tau+x)
   \nonumber\\
   &\mathrel{\phantom{=}}
   \times
   \ell_{\gamma}\!\left(\frac{t-\tau+x}{2u^{1/\gamma}}\right)
   \ell_{\gamma}\!\left(\frac{t-\tau-x}{2u^{1/\gamma}}\right)
   \nonumber\\
   &\mathrel{\phantom{=}}
   \times
   \ell_{\gamma}\!\left(\frac{\tau+y}{2u^{-1/\gamma}}\right)
    \ell_{\gamma}\!\left(\frac{\tau-y}{2u^{-1/\gamma}}\right).
    \label{F_t(x,y)2}
\end{align}

Finally, by changing variables, $\tau=(t-|x|-|y|)\eta+|y|$ for the internal integral in  Eq.~(\ref{F_t(x,y)2}) and $u=\left(\frac{t-|x|-|y|}{2}\right)^{\gamma}\!\vartheta$ for the external one, from
$P_{XY}(x,y,t)=H(x,y,t)+H(-x,y,t)+H(y,x,t)+H(-y,x,t)$ [see Eq.~\eqref{P_ks1}]
we obtain
%\begin{widetext}
%\begin{eqnarray}
%   P_{XY}(x,y,t) \!\!&=&\!\!\frac{4 t}{\gamma(t-|x|-|y|)^3} \int_{0}^{\infty}\mathrm{d}\vartheta\,
%   \vartheta^{-4/\gamma-1}\! \int_{0}^{1}\mathrm{d}\eta\,
%   \ell_{\gamma}\!\left[\vartheta^{-1/\gamma}(1-\eta)\right]
%   \ell_{\gamma}\!\left(\vartheta^{-1/\gamma} \eta\right)
%   \nonumber\\[4pt]
%   &&\!\!\times
%   \ell_{\gamma}\!\left[\vartheta^{-1/\gamma}\!\left(\!1-\eta+x_t\right) \right]
%   \ell_{\gamma}\!\left[\vartheta^{-1/\gamma}\!\left(\eta+y_t\right) \right]
%    \label{P_t(x,y)3}
%\end{eqnarray}
%\end{widetext}
\begin{align}
   P_{XY}(x,y,t) &= \frac{4 t}{\gamma(t-|x|-|y|)^3} \int_{0}^{\infty}\mathrm{d}\vartheta\,
   \vartheta^{-4/\gamma-1}
   \nonumber\\
   &\mathrel{\phantom{=}}
   \times\int_{0}^{1}\mathrm{d}\eta\,
   \ell_{\gamma}\!\left(\frac{1-\eta}{\vartheta^{1/\gamma}}\right)
   \ell_{\gamma}\!\left( \frac{\eta}{\vartheta^{1/\gamma}}\right)
   \nonumber\\
   &\mathrel{\phantom{=}}
   \times
   \ell_{\gamma}\!\left(\frac{1-\eta+x_t}{\vartheta^{1/\gamma}}\right)
   \ell_{\gamma}\!\left(\frac{\eta+y_t}{\vartheta^{1/\gamma}}\right)
    \label{P_t(x,y)3}
\end{align}
if $|x|+|y|<t$ and $P_{XY}(x,y,t)=0$ if otherwise.

Expression (\ref{P_t(x,y)3}) is less complex than
the one obtained in the previous section but it includes a double integral and seems to be  less convenient for numerical evaluation. However, as we will show in Section~\ref{sec:6}, this form allows us to derive an exact analytic expression for the case $\gamma=1/2$.
Moreover, from this representation we see that $P_{XY}(x,y,t)$
is indeed a non-negative function and hence it is a legitimate pdf [while the normalization condition is obviously holds  due to the fact that $P_{XY}(\mathbf{k},s)|_{\mathbf{k}=0}=1/s$].

%\begin{comment}
By changing variables in Eq.~(\ref{P_t(x,y)3}),  $\vartheta^{-1/\gamma}\mapsto\vartheta$, $\overline{x}=x/t$,
and $\overline{y}=y/t$, and introducing new variables,
\begin{equation}
    x_r=\frac{2|\overline{x}|}{1-|\overline{x}|-|\overline{y}|},
    \quad y_r=\frac{2|\overline{y}|}{1-|\overline{x}|-|\overline{y}|},
    \label{xr_yr}
\end{equation}
we obtain the following expression for   $\mathcal{P}(\overline{x},\overline{y})$:
\begin{align}
   \mathcal{P}(\overline{x},\overline{y}) &=  \frac{4}{(1-|\overline{x}|-|\overline{y}|)^3} \int_{0}^{\infty}\mathrm{d}\vartheta
   \vartheta^3 \int_{0}^{1}\mathrm{d}\eta\, \ell_{\gamma}\!\left[\vartheta(1-\eta)\right]
   \nonumber\\[2pt]
   &\mathrel{\phantom{=}}
   \times
   \ell_{\gamma}\!\left(\vartheta\eta \right)
   \ell_{\gamma}\!\left[\vartheta\left(1-\eta+x_r\right) \right]
   \ell_{\gamma}\!\left[\vartheta\!\left(\eta+y_r\right) \right]
    \label{P_t(x,y)4}
\end{align}
when $|\overline{x}|+|\overline{y}|<1$ and  $\mathcal{P}(\overline{x},\overline{y})=0$ otherwise \cite{comment0}.

\section{Numerical analysis}
\label{sec:5}

\subsection{Numerical evaluation of $\mathcal{P}(\overline{x},\overline{y})$}

Here we show how to compute  asymptotic pdf
$\mathcal{P}(\overline{x},\overline{y})$.
%by using Eqs.~(\ref{P_XY_Q12} -\ref{Q2x}).
\begin{figure*}[t]
   \includegraphics[width=1\textwidth]{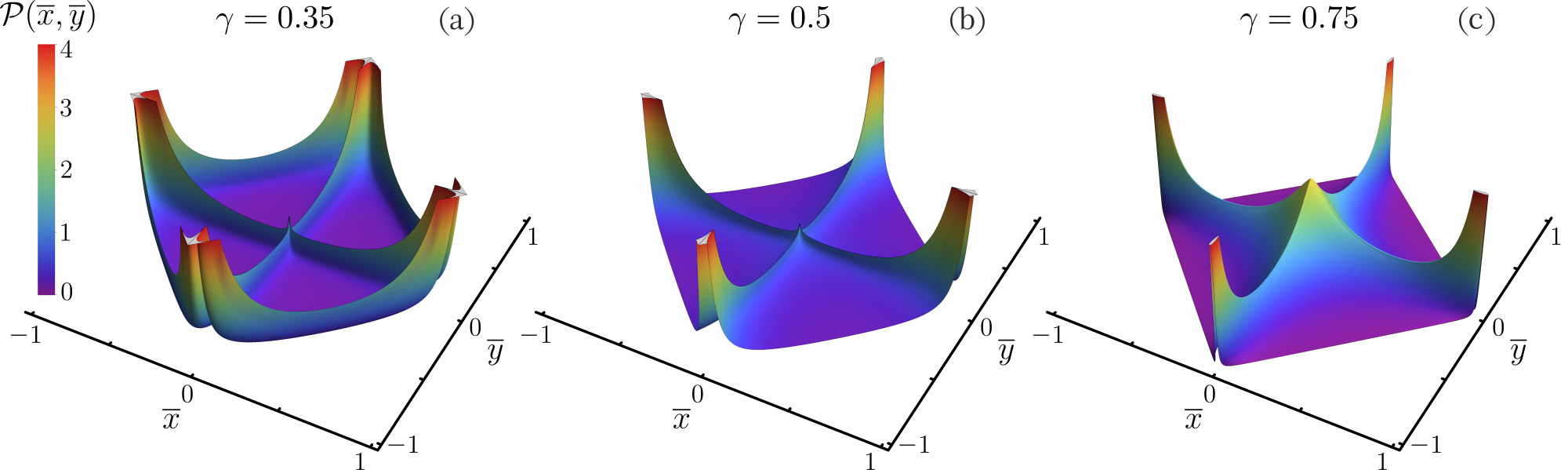}
    \caption{Asymptotic probability density functions  $\mathcal{P}(\overline{x},\overline{y})$ for different values of $\gamma$. The functions are obtained by using Eq.~\eqref{P_num}. Note that in cases (a) and (b) $\mathcal{P}(\overline{x},\overline{y})$ is singular along lines $\overline{x}=0$ and $\overline{y}=0$,
    %therefore points with these coordinates are not taken into a consideration
    while minimal values of $\overline{x}$ and $\overline{y}$ used to plot the functions are $10^{-3}$. In the case (a), $\mathcal{P}(\overline{x},\overline{y})$ is also singular along the ballistic front   $|x|+|y|=1$ and the outward points used to plot the functions are at distance $10^{-3}$ from the front.}
    \label{fig:2}
\end{figure*}
From Eq.~\eqref{P_XY_Q12} we have
\begin{equation}
    \mathcal{P}(\overline{x},\overline{y}) =  Q_1(\overline{x},\overline{y})+ Q_2(\overline{x},\overline{y})+Q_1(\overline{y},\overline{x})+Q_2(\overline{y},\overline{x}),
    \label{P_num}
\end{equation}
where  $Q_{1,2}(\overline{x},\overline{y})=Q_{1,2}(\overline{x},\overline{y},t=1)$. In Eqs.~\eqref{Q1x} and \eqref{Q2x} we replace variable $\eta\mapsto\frac{1+\eta}{2}$. This allows us to extend the integration interval from $[0,1]$ to $[-1,1]$ and then implement Gauss–Jacobi quadrature \cite{RalstonRabinowitz}. We chose this particular method because it is very convenient to deal numerically with integrals which includes power-law singularities.

We now write
\begin{align}
    Q_1(\overline{x},\overline{y}) &= \int_{-1}^{1}\!\mathrm{d}\eta \, (1-\eta)^{\gamma-1}q_1(\eta;\overline{x},\overline{y}),
    \label{Q12x}
    \\
     Q_2(\overline{x},\overline{y}) &= \int_{-1}^{1}\!\mathrm{d}\eta \, (1-\eta)^{\gamma-1}q_2(\eta;\overline{x},\overline{y})
    \label{Q12x1}
\end{align}

with functions
\begin{widetext}
\begin{align}
    q_1(\eta;\overline{x},\overline{y}) &=
    \frac{2(1-\gamma)}{\pi^2 x_r^{\gamma-1} y_r (1-|\overline{x}|-|\overline{y}|)^2} (1-\eta)^{1-\gamma} \left(1-\eta+2x_r\right)^{\gamma-2}
    \nonumber\\
    &\mathrel{\phantom{=}}
    \times\text{Re} \left\{
    \frac{1}{x_r^{-\gamma}\left[(1-\eta)^{\gamma}e^{i\pi\gamma}+(1-\eta+2x_r)^{\gamma}\right] +y_r^{-\gamma}\left[(1+\eta)^{\gamma}e^{-i\pi\gamma}+(1+\eta+2y_r)^{\gamma}\right]}
    \right.
    \nonumber\\
    &\mathrel{\phantom{=}}
    -\left.
     \frac{1}{x_r^{-\gamma}\left[(1-\eta)^{\gamma}e^{i\pi\gamma}+(1-\eta+2x_r)^{\gamma}\right] +y_r^{-\gamma}\left[(1+\eta)^{\gamma}e^{i\pi\gamma}+(1+\eta+2y_r)^{\gamma}\right]}
    \right\},
    \label{q1x}
\end{align}
\begin{align}
   q_2(\eta;\overline{x},\overline{y}) &=
    \frac{2\gamma}{\pi^2 x_r^{\gamma}y_r(1-|\overline{x}|-|\overline{y}|)^2}
   \left[x_r^{1-\gamma}\left(1-\eta+2x_r\right)^{\gamma-1} +y_r^{1-\gamma}\left(1+\eta+2y_r\right)^{\gamma-1}\right]
    \nonumber\\
    &\mathrel{\phantom{=}}
    \times
   \text{Re} \left[e^{i\pi(\gamma-1)}+(1-\eta)^{1-\gamma}(1-\eta+2 x_r)^{\gamma-1}\right]
    \nonumber\\
    &\mathrel{\phantom{=}}
    \times \left\{
    \frac{1}{\left\{x_r^{-\gamma}\left[(1-\eta)^{\gamma}e^{i\pi\gamma}+(1-\eta+2x_r)^{\gamma}\right] +y_r^{-\gamma}\left[(1+\eta)^{\gamma}e^{-i\pi\gamma}+(1+\eta+2y_r)^{\gamma}\right]\right\}^2}
    \right.
    \nonumber\\
    &\mathrel{\phantom{=}}
    -\left.
     \frac{1}{\left\{x_r^{-\gamma}\left[(1-\eta)^{\gamma}e^{i\pi\gamma}+(1-\eta+2x_r)^{\gamma}\right] +y_r^{-\gamma}\left[(1+\eta)^{\gamma}e^{i\pi\gamma}+(1+\eta+2y_r)^{\gamma}\right]\right\}^2}
    \right\},
    \label{q2x}
\end{align}
\end{widetext}
which have no singularities with respect to $\eta$ (for any fixed values of  $\overline{x}$ and $\overline{y}$).

Following the Gauss–Jacobi quadrature recipe \cite{RalstonRabinowitz}, we obtain
\begin{align}
    Q_1(\overline{x},\overline{y}) &\approx \sum_{j=1}^{n} w_j q_1(\eta_j;\overline{x},\overline{y}),
    \label{Q12x_approx}
    %\nonumber
    \\
     Q_2(\overline{x},\overline{y}) &\approx \sum_{j=1}^{n} w_j q_2(\eta_j;\overline{x},\overline{y}),
    \label{Q12x_approx1}
    %\nonumber
\end{align}
where weights are
\begin{align}
    w_j &=-\frac{(2n+a+b+2)\Gamma(n+a+1)}{(n+a+b+1)^2\Gamma(n+a+b+1)}
    \nonumber\\
    &\mathrel{\phantom{=}}
    \times\frac{\Gamma(n+b+1)2^{a+b+1}}{\Gamma(n+2)J_{n-1}^{(a+1,b+1)}(\eta_j)J_{n+1}^{(a,b)}(\eta_j)}
    \label{w_j}
    %\nonumber
\end{align}
and $\eta_j$ are roots of Jacobi polynomials  $J_{n}^{(a,b)}(\eta)$. In our case $a=\gamma-1$ and $b=0$.

\begin{figure*}[t]
	\includegraphics[width=0.7\textwidth]{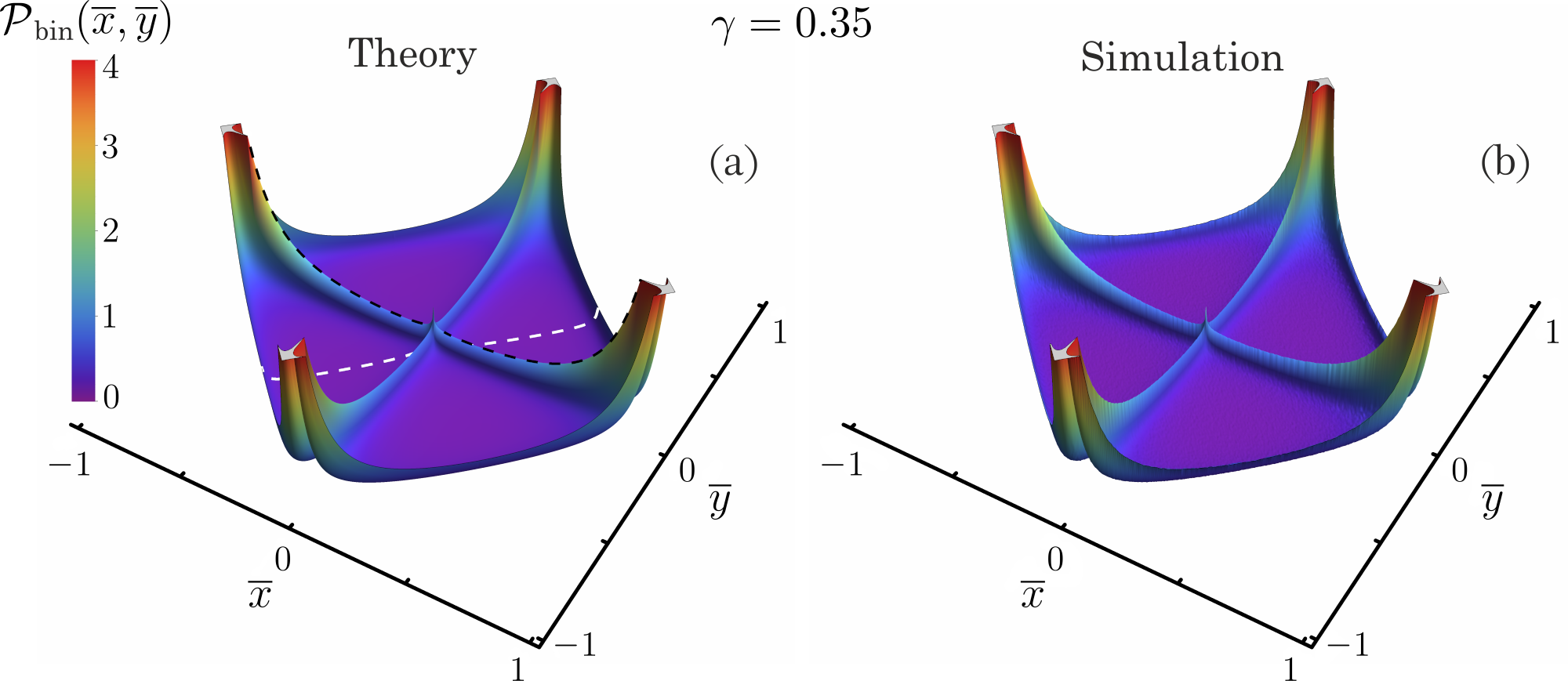}
    \caption{Averaged probability density functions $\mathcal{P}_{\mathrm{bin}}(\overline{x},\overline{y})$ for $\gamma=0.35$ obtained  (a) with Eq.~\eqref{P_Aij} and (b) by sampling a histogram for $t=10^3$ with $10^8$ realizations. To calculates the functions, the square $[-1,1] \times [-1,1]$ was divided into a grid of $400\times400$ with cells. Sections $\overline{y}=0$ (black dashed line) and $\overline{y}=\overline{x}$ (white dashed line)
    are presented on Figures~5(a) and 6(a), respectively. Number of the sampled realisations is $10^8$.}
	\label{fig:3}
\end{figure*}

\begin{figure*}[htp]
	\includegraphics[width=0.7\textwidth]{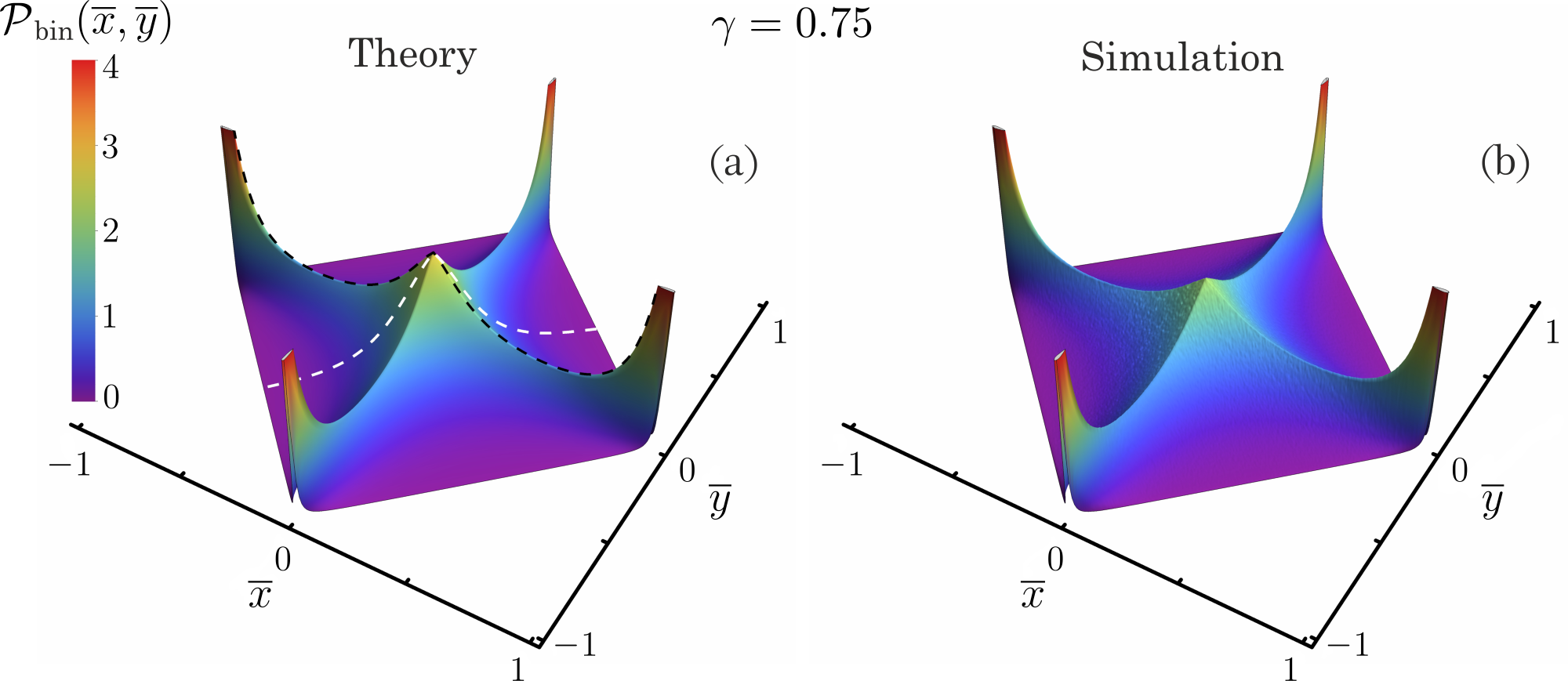}
     \caption{Averaged probability density functions $\mathcal{P}_{\mathrm{bin}}(\overline{x},\overline{y})$ for $\gamma=0.75$ obtained  (a) with Eq.~\eqref{P_Aij} and (b) by sampling a histogram for $t=10^3$ with $10^8$ realizations. To calculates the functions, the square $[-1,1] \times [-1,1]$ was divided into a grid of $400\times400$ bins. Distributions along the sections $\overline{y}=0$ (black dashed line) and $\overline{y}=\overline{x}$ (white dashed line)
    are presented on Figures~5(c) and 6(c), respectively. Number of the sampled realisations is $10^8$.}
	\label{fig:4}
\end{figure*}

In  the functions under  the integrals in Eqs.~\eqref{Q12x} and \eqref{Q12x1},
we separate singular multiplier $(1-\eta)^{\gamma-1}$, and then compensate it with  $(1-\eta)^{1-\gamma}$ in some places \cite{comment1}. Figure~\ref{fig:2} shows asymptotic pdf computed for three different values of $\gamma$. Note that, for $\gamma=0.35$ and $0.5$, the corresponding pdf's are singular along lines $\overline{x}=0$ and $\overline{y}=0$ \cite{to_be_published}. Additionally, for $\gamma=0.35$, the pdf is also singular along the ballistic front \cite{to_be_published}.
The numerically calculated pdf's have finite height because the minimal distances of the grid points from the singular lines are $10^{-3}$.

\subsection{Comparison with the results of finite-time samplings}
%average probability density over a bin  $\mathcal{P}_{\text{bin}}(\overline{x},\overline{y})$}
Here we discuss a procedure to compare analytical results with numerically sampled finite-time histograms.

%In order to compare the  analytical results, obtained for  %$\mathcal{P}(\overline{x},\overline{y})$,
%with the numerically sampled histograms,
We split the domain where the pdf takes non-zero values, i.e. inside the ballistic square $|\overline{x}|+|\overline{y}|<1$, into a set of bins.

\begin{figure*}[t]
   \includegraphics[width=0.7\textwidth]{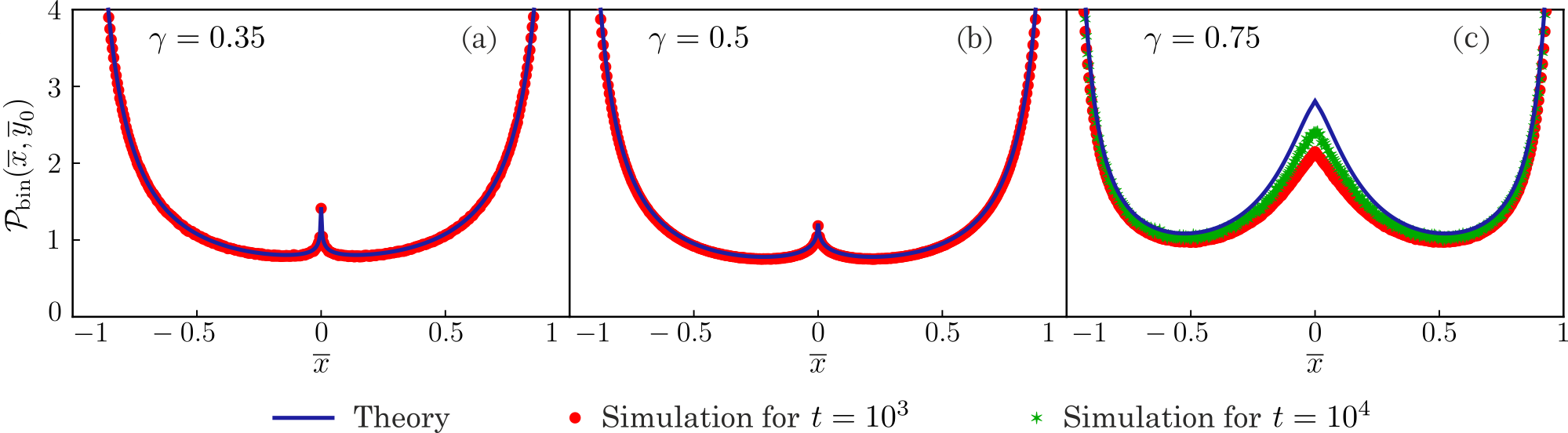}
    \caption{Sections of $\mathcal{P}_{\mathrm{bin}}(\overline{x},\overline{y})$ along line $\overline{y}=\overline{y}_0 = 0$ for three different values of $\gamma$. Blue solid curves are  theoretic results, red circles are result of the sampling for time $t=10^3$, and green stars [on panel (c)] are result of the sampling  for time $t=10^4$. Number of the sampled realisations is $10^8$ in all the cases.}
    \label{fig:5}
\end{figure*}

\begin{figure*}[htp]
   \includegraphics[width=0.7\textwidth]{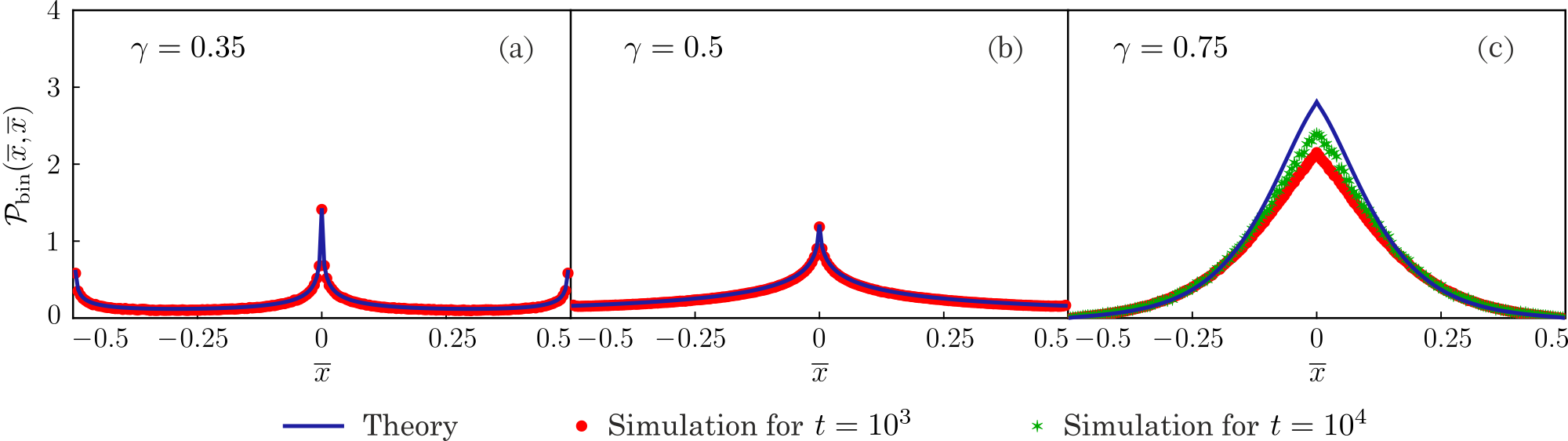}
    \caption{Sections of $\mathcal{P}_{\mathrm{bin}}(\overline{x},\overline{y})$  along line $\overline{y}=\overline{x}$ for three different values of $\gamma$. Blue solid curves are  theoretic results, red circles are result of the sampling for time $t=10^3$, and green stars [on panel (c)] are result of the sampling  for time $t=10^4$. Number of the sampled realisations is $10^8$ in all the cases.}
    \label{fig:6}
\end{figure*}

Consider now bin  $\mathcal{A}$ of area $S(\mathcal{A})$. Then the average probability density function over bin  $\mathcal{A}$ is
\begin{equation}
    \mathcal{P}_{\mathcal{A}}=\frac{1}{S(\mathcal{A})}\iint_{\mathcal{A}}\mathcal{P}(\overline{x},\overline{y})    \mathrm{d}\overline{x}\mathrm{d}\overline{y}.
    \label{P_A}
\end{equation}
The corresponding pdf (which will be estimated through the numerical sampling) is
\begin{equation}
    \mathcal{P}^{\mathrm{num}}_{\mathcal{A}}=\frac{1}{S(\mathcal{A})}\frac{N_{\mathcal{A}}}{N_{\mathrm{total}}}.
\end{equation}
Here
$N_{\mathcal{A}}$ is the number of realizations which ended up, after fixed time $t$, in bin $\mathcal{A}$, while
$N_{\mathrm{total}}$ is the total number of realizations.
Then, for large enough $N_{\mathrm{total}}$, we expect $\mathcal{P}_{\mathcal{A}} \approx \mathcal{P}^{\mathrm{num}}_{\mathcal{A}}$.

If function $\mathcal{P}(\overline{x},\overline{y})$ is continuous over $\mathcal{A}$, then, according to the mean value theorem,
there is point  $(\overline{x}_c,\overline{y}_c)\in\mathcal{A}$ such that $\mathcal{P}_{\mathcal{A}}=\mathcal{P}(\overline{x}_c,\overline{y}_c)$.
If, in addition, $\mathcal{A}$ is small (compared to the characteristic scale over which $\mathcal{P}_{\mathcal{A}}$ varies substantially), then, by using Taylor series,  we have  $\mathcal{P}(\overline{x},\overline{y})\approx  \mathcal{P}(\overline{x}_c,\overline{y}_c)$ for all  $(\overline{x},\overline{y})\in\mathcal{A}$. Therefore, in  a sufficiently small  domain $\mathcal{A}$,
pdf $\mathcal{P}(\overline{x},\overline{y})$  can be approximated by the average density over the domain such that $\mathcal{P}(\overline{x},\overline{y})|_{(\overline{x},\overline{y})\in\mathcal{A}}\approx \mathcal{P}_{\mathcal{A}} \approx \mathcal{P}^{\mathrm{num}}_{\mathcal{A}}$.

We partition the interior of square $|\overline{x}|+|\overline{y}|<1$ into set of bins with a set of lines parallel to the main Cartesian  axes and distance  $\varepsilon$ between two neighboring lines. By doing that, we obtain  $M^2$ bins,  $M = 1/\varepsilon$.

%Note that counting should be performed when the partitions is implemented by using the operation of taking integer part of the coordinates. Namely, the bin around the origin gets approximately  four times more realizations that a non-specific a bin, while a bin along one of the axes could get two times  more.

Bin $\mathcal{A}_{i j}$ is defined as $\overline{x}_i\leq \overline{x}\leq \overline{x}_{i+1}$ and $\overline{y}_j\leq \overline{y}\leq \overline{y}_{j+1}$ with $\overline{x}_{i+1}-\overline{x}_i=\overline{y}_{j+1}-\overline{y}_j=\varepsilon$.
We have
\begin{align}
    \mathcal{P}_{\mathcal{A}_{i j}} &= \frac{1}{S(\mathcal{A}_{i j})}\iint_{\mathcal{A}_{i j}}\mathcal{P}(\overline{x},\overline{y})    \mathrm{d}\overline{x}\mathrm{d}\overline{y}
    \nonumber\\
    &=\frac{1}{\varepsilon^2}\int_{\overline{x}_{i}}^{\overline{x}_{i+1}}\!\int_{\overline{y}_{j}}^{\overline{y}_{j+1}}
     \mathcal{P}\!\left(\overline{x},\overline{y}\right)\mathrm{d}\overline{x}\mathrm{d}\overline{y}.
     \nonumber
\end{align}
By introducing variables $\overline{x}=\frac{\overline{x}_{i+1}-\overline{x}_{i}}{2}x^{\prime}+\frac{\overline{x}_{i+1}+\overline{x}_{i}}{2}$
and $\overline{y}=\frac{\overline{y}_{j+1}-\overline{y}_{j}}{2}y^{\prime}+\frac{\overline{y}_{j+1}+\overline{y}_{j}}{2}$,  which maps intervals $[\overline{x}_{i}, \overline{x}_{i+1}]$ and $[\overline{y}_{j}, \overline{y}_{j+1}]$ onto the interval $[-1,1]$, we get
\begin{widetext}
\begin{align}
     \mathcal{P}_{\mathcal{A}_{i j}} &= \frac{1}{4} \int_{-1}^{1}\!\int_{-1}^{1}
      \mathcal{P}\!\left[\frac{\varepsilon}{2} (x^{\prime}+1)+\overline{x}_i,
      \frac{\varepsilon}{2} (y^{\prime}+1)+\overline{y}_j\right]
      \mathrm{d}x^{\prime}\mathrm{d}y^{\prime}
      \nonumber\\
      &
      \approx
       \frac{1}{4}\sum_{m_{1}=1}^{m} \sum_{m_{2}=1}^{m} w_{m_1}w_{m_2}
      \mathcal{P}\!\left[\frac{\varepsilon}{2} (x^{\prime}_{m_1}+1)+\overline{x}_i,
      \frac{\varepsilon}{2} (y^{\prime}_{m_2}+1)+\overline{y}_j\right]\!.
      \label{P_Aij}
\end{align}
\end{widetext}

In the second line of Eq.~\eqref{P_Aij} we approximate the integral by using orthogonal Legendre polynomials  of order $m$ (they can be obtained as a particular case of Jacobi polynomials by setting $a=b=0$). Therefore, in this case weights $w_{m_1,m_2}$ follows  from Eq.~\eqref{w_j} with $a=b=0$, while $x_{m_1}$ and $y_{m_2}$ are root of Legendre polynomial $P_m(\xi)=J_{m}^{(0,0)}(\xi)$.  To find $\mathcal{P}_{\mathcal{A}_{i j}}$, we have to calculate
$ \mathcal{P}\!\left[\frac{\varepsilon}{2} (x^{\prime}_{m1}+1)+\overline{x}_i, \frac{\varepsilon}{2} (y^{\prime}_{m2}+1)+\overline{y}_j\right]$
by using the above described method. For the bins   ${A}_{i j}$, in which  $\mathcal{P}(\overline{x},\overline{y})$ has singularities \cite{to_be_published}, we can use the same scheme, by taking into account the corresponding singularity order in Eq.~\eqref{P_Aij}.

We denote the averaged (over the bin) probability density  $\mathcal{P}_{\mathcal{A}_{i j}}$ as  $\mathcal{P}_{\mathrm{bin}}(\overline{x},\overline{y})$, where
$(\overline{x},\overline{y})$ are coordinates of the center of the corresponding bin  $\mathcal{A}_{i j}$. To compare   $\mathcal{P}_{\mathrm{bin}}(\overline{x},\overline{y})$ with $\mathcal{P}(\overline{x},\overline{y})$, we calculate pdf $\mathcal{P}(\overline{x},\overline{y})$ at the center of the corresponding bin $\mathcal{A}_{i j}$. 
%Note that an outer bin gets only trajectories which a limited by the ballistic front and therefore end up in the first half of the bin; in the case of a corner bin it is a quarter of its area. The underline here is that  a proper normalization has to be performed.

Figures \ref{fig:4}--\ref{fig:6} present a comparison of the probability distributions obtained by averaging pdf $\mathcal{P}(\overline{x},\overline{y})$, Eq.~(\ref{P_scale}),
over the bins of a $400\times400$ grid, with the results of a finite-time sampling (by using the same grid) with $10^8$ realizations. While for $\gamma = 0.35$ and $0.5$ sampling over time $t=10^3$ yields histograms that are in a perfect  agreement with the theoretical results, for $\gamma = 0.75$ the peak  at the origin  develops slowly in time.

\section{Exact solution for $\gamma=1/2$}
\label{sec:6}

%Here we derive an exact expression for  $\mathcal{P}(\overline{x},\overline{y})$ for a particular value of $\gamma$.

For  $\gamma=1/2$  we have L\'{e}vy-Smirnov
distribution \cite{SamorodnitskyTaqqu}
\begin{equation}
    \ell_{1/2}(t)=\frac{e^{-1/(4t)}}{2\sqrt{\pi}t^{3/2}}.
    \label{LS}
    \nonumber
\end{equation}
By substituting this into Eq.~\eqref{P_t(x,y)4}, after some elementary calculations, we obtain
\begin{equation}
   \mathcal{P}(\overline{x},\overline{y}) =  \frac{4}{\pi^2 (1-|\overline{x}|-|\overline{y}|)^3}\int_{0}^{1}\frac{\mathrm{d}\eta\sqrt{p(\eta)}}
   {\widetilde{p}^2(\eta)},
    \label{P_1/2}
    \nonumber
\end{equation}
where
\begin{equation}
    p(\eta)=(1+x_r-\eta)(1-\eta)\eta(\eta+y_r)
    \label{p}
\end{equation}
and
\begin{equation}
    \widetilde{p}(\eta)=p(\eta)\left(\frac{1}{1+x_r-\eta}+\frac{1}{1-\eta}+\frac{1}{\eta}+\frac{1}{\eta+y_r}\right).
    \label{p_tilde}
    \nonumber
\end{equation}

Next, we take into account that
\begin{equation}
    \widetilde{p}(\eta)=-(2+x_r+y_r)(\eta-\eta_1)(\eta-\eta_2),
    \label{znam}
    \nonumber
\end{equation}
where
\begin{equation}
   \eta_{1,2}=\frac{1+x_r\pm\sqrt{(1+x_r)(1+y_r)(1+x_r+y_r)}}{2+x_r+y_r}.
    \label{eta12}
\end{equation}
Since $1-|\overline{x}|-|\overline{y}|=\frac{2}{2+x_r+y_r}$ [as it follows from Eq.~\eqref{xr_yr}], we can recast the expression for the asymptotic pdf in the form
\begin{equation}
   \mathcal{P}(\overline{x},\overline{y}) = \frac{2+x_r+y_r}{2 \pi^2 } \int_{0}^{1} \frac{\mathrm{d}\eta \sqrt{p(\eta)}}{(\eta-\eta_1)^2(\eta-\eta_2)^2}.
    \label{P_1/2_1}
\end{equation}

Because $p(\eta)$ is polynomial of the fourth order,  the integral in Eq.~(\ref{P_1/2_1})  can be expressed through elliptic integrals \cite{Hancock} (see Appendix \ref{sec:AppB} for more details):
\begin{align}
   \mathcal{P}(\overline{x},\overline{y}) &= \Omega(\overline{x},\overline{y})\!\left[ \left(\mu(\overline{x},\overline{y})+\frac{1}{\mu(\overline{x},\overline{y})}\right)
   K\!\left(1-\frac{1}{\mu^2(\overline{x},\overline{y})}\right)\right.
   \nonumber\\
   &\mathrel{\phantom{=}}
   \left.-2\mu(\overline{x},\overline{y}) E\!\left(1-\frac{1}{\mu^2(\overline{x},\overline{y})}\right)\right],
    \label{P_1/2_2}
\end{align}
where
\begin{align}
   \Omega(\overline{x},\overline{y}) &= \frac{(2+x_r+y_r)^3}{16\pi^2}
    \nonumber\\
   &\mathrel{\phantom{=}}
   \times \frac{\sqrt{(1+x_r)(1+y_r)}-\sqrt{1+x_r+y_r}} {(1+x_r)(1+y_r)(1+x_r+y_r)},
    \label{P_1/2_3}
\end{align}
\begin{equation}
     \mu(\overline{x},\overline{y})=
     \frac{\sqrt{(1+x_r)(1+y_r)}+\sqrt{1+x_r+y_r}}{\sqrt{(1+x_r)(1+y_r)}-\sqrt{1+x_r+y_r}},
    \label{mu}
\end{equation}
while  $K(m)$ and $E(m)$ complete elliptic integrals of the first and second kind.

\section{Conclusions}
\label{sec:7}

In this work we present a detailed theoretical analysis of a particular two-dimensional L\'{e}vy walk (LW) model in the ballistic regime. With this, we wanted to demonstrate that a complex planar spatially anisotropic LW process~\cite{prl2016} can be evaluated analytically up to fine details \cite{asymptotics}. In this context, our work constitutes a next step in the direction set in Refs.~\cite{zarfaty1,zarfaty2}, where this program was realized for  the border between diffusive  and superdiffusive regimes, i.e., for $\gamma = 2$.

The super-diffusive regime, $1 < \gamma < 2$, was partially addressed in Ref.~\cite{fouxon2017}. This regime, however, is the hardest one to evaluate analytically. In this case $P(x,y,t)$  does not obey a uniform scaling  but rather two different ones, a L\'{e}vy scaling governing the bulk of the pdf and 
co-variant scaling \cite{covariant} governing the ballistic ends. In $2d$, aside of the obvious dependence of the  scalings on the direction,  the position of the 'meeting' point of the two scalings depends not only on time (as in the one-dimensional case \cite{covariant}) but also on the direction. We hope that a progress will be made in this  direction and it would be possible, f.e., to relate  analytical  considerations and  numerical simulations of transport processes in two-dimensional Hamiltonian systems \cite{yosi3}.

\section{Acknowledgments}\label{acknowledgment} S.D. appreciate the hospitality of the
Max Planck Institute for the Physics of Complex Systems (Dresden, Germany)  where the project was finalized. He also acknowledges support by the Nord-STAR - Nordic Center for Sustainable and Trustworthy AI Research (OsloMet Project Code 202237-100).

%%%%%%%%%%%%%%%%%%%%%%%%%%%%%%%%%%%%%%%%%%%%%%%%%%%%%%%%%%%%%
\appendix

\section{Inverse Fourier-Laplace transform in 1d case}
\label{sec:AppA}
\setcounter{figure}{0}

Here we briefly review the method presented by  Godr\`{e}che and Luck in Ref.~\cite{GL2001}.

Assume there is scaling  $G(x,t)=\frac{1}{t}\Phi\!\left(\frac{x}{t}\right)$. We denote   $\overline{x}=\frac{x}{t}$ and obtain
\begin{align}
     G(k,s)&=\mathcal{F}_x\mathcal{L}\{G(x,t)\}=\int_{-\infty}^{\infty}\frac{\Phi(\overline{x})\mathrm{d}\overline{x}}{i k \overline{x}+s }
     \nonumber\\
     &=
     \frac{1}{s}\left\langle \frac{1}{1+\frac{i k}{s}\overline{X}}\right\rangle=\frac{1}{s}g\!\left(\frac{i k}{s}\right).
     \label{G_ks}
\end{align}
According to the Sokhotski–Plemelj theorem \cite{Vladimirov}
\begin{equation}
    \lim\limits_{\,\epsilon \rightarrow 0^+}\frac{1}{\xi\pm i\epsilon}=\mp i\pi \delta(\xi)+\mathscr{P}\frac{1}{\xi}
     %\label{SP}
     \nonumber
\end{equation}
(a letter $\mathscr{P}$ denotes that the Cauchy  principal value is taken) and, therefore, $\delta(\xi)\!=\!-\frac{1}{\pi}\text{Im}\!\lim\limits_{\epsilon \rightarrow 0^+}\frac{1}{\xi+ i\epsilon}$.

Then, taking into account that
\begin{equation}
    \Phi(\overline{x})=\langle\delta(\overline{x}-\overline{X})\rangle,
    \label{Phi_def}
\end{equation}
we obtain
\begin{align}
     \Phi(\overline{x}) &=  -\frac{1}{\pi}\text{Im}\!\lim\limits_{\,\epsilon \rightarrow 0^+}\left\langle\frac{1}{\overline{x}-\overline{X} + i\epsilon}\right\rangle
     \nonumber\\
    % &=
    %-\frac{1}{\pi}\!\lim\limits_{\,\epsilon \rightarrow 0^+}\text{Im}\!\left[
    %\frac{1}{\overline{x}+ i\epsilon}\left\langle\frac{1}{1-\frac{1}{\overline{x}+ i\epsilon}\overline{X}}\right\rangle\right]
     %\nonumber\\
     &=
     -\frac{1}{\pi}\!\lim\limits_{\,\epsilon \rightarrow 0^+}\text{Im}\!\left[
    \frac{1}{\overline{x}+ i\epsilon} g\!\left(-\frac{1}{\overline{x}+ i\epsilon}\right) \right].
     \label{Phi}
\end{align}
Therefore
\begin{eqnarray}
     G(x,t) = -\frac{1}{\pi x}\!\lim\limits_{\,\epsilon \rightarrow 0^+}\!\!\text{Im}\, g\!\left(-\frac{1}{x/t+ i\epsilon}\right).
     \label{G(x,t)_app}
\end{eqnarray}

Next we show that the method proposed by Godr\`{e}che and  Luck is related to the Stieltjes transform.

Namely, from Eq.~(\ref{G_ks}) follows (here we introduce notation $i k/s=\zeta$)
\begin{equation}
     g(\zeta)=\int_{-\infty}^{\infty}\frac{\Phi(\overline{x})\mathrm{d}\overline{x}}{1+\zeta\,\overline{x}}.
     %\label{g_zeta}
     \nonumber
\end{equation}
We make replace $\zeta=-1/z$ and obtain
\begin{equation}
     \frac{1}{z}g\!\left(-\frac{1}{z}\right)=\int_{-\infty}^{\infty}\frac{\Phi(\overline{x})\mathrm{d}\overline{x}}{z-\overline{x}}.
     \label{g_zeta1}
\end{equation}
The rhs of Eq.~(\ref{g_zeta1}) is the Stieltjes transform of $\Phi(\overline{x})$.
Denoting  $\mathcal{S}(z)=\frac{1}{z}g\!\left(-\frac{1}{z}\right)$, taking into account that the inverse  Stieltjes transform is defined as \cite{Widder}
\begin{equation}
     \Phi(\overline{x})=\lim\limits_{\,\epsilon \rightarrow 0^+}\! \frac{\mathcal{S}(\overline{x}-i\epsilon)-\mathcal{S}(\overline{x}+i\epsilon)}{2\pi i},
     %\label{Phi1}
     \nonumber
\end{equation}
and using the identity  $\text{Im}\,f=\frac{f-f^*}{2 i}$, we arrive at the formula~(\ref{Phi}).

Therefore, it is clear now that the method, in principle, is a particular case of the implementation of the inverse  Stieltjes transform. Usually it is used for the probability density functions. However, it is not specific and can be used for general functions \cite{Widder}, like functions in Eqs.~(\ref{G_1}) and (\ref{G_2}).

\section{Legendre's normal  form for elliptic integrals}
\label{sec:AppB}

Assume that  $R(\eta)$ is a forth-order polynomial and  $S(\eta)$ is an arbitrary rational function. We will follow Ref.~\cite{Hancock} (Section VIII in there), and describe a method to reduce integrals of the following type
\begin{equation}
   I=\int_0^1\frac{S(\eta)\mathrm{d}\eta}{\sqrt{R(\eta)}}
    \label{Leg}
\end{equation}
to elliptic ones. We are only interested in the case when
all roots of $R(\eta)$ are real; we also set the leading coefficient of the polynomial equals to one. We write $R(\eta)=(\eta-a_1)(\eta-a_2)(\eta-a_3)(\eta-a_4)$ and apply to Eq.~(\ref{Leg}) the following linear fractional transform
\begin{equation}
   \eta=\frac{a\omega+b}{c\omega+d},
    \label{frac_lin}
\end{equation}
where we also assume  $a d - b c\neq 0$. Next we take into account that
\begin{equation}
   \eta-a_j=\frac{(a-c a_j)\omega+b-d a_j}{c\omega+d}
    %\label{frac_lin1}
    \nonumber
\end{equation}
with $j=\overline{1,4}$ and
\begin{equation}
   \mathrm{d}\eta=\frac{a d - b c}{(c\omega+d)^2}\mathrm{d}\omega,
    %\label{frac_lin2}
    \nonumber
\end{equation}
and arrive at
\begin{equation}
   I=\int_{-\frac{b}{a}}^{-\frac{b-d}{a-c}}
   \frac{(a d - b c)\sigma(\omega)\mathrm{d}\omega}
   {\sqrt{\prod\limits_{j=1}^{4}[(a-c a_j)\omega+b-d a_j]}},
    \label{Leg2}
\end{equation}
where $\sigma(\omega)=S\!\left(\frac{a\omega+b}{c\omega+d}\right)$ is a rational function.

Next we write
\begin{align}
  \prod\limits_{j=1}^{4}[(a-c a_j)\omega+b-d a_j]&=(q_0\omega^2+q_1\omega+q_2)
   \nonumber\\
    &\mathrel{\phantom{=}}
    \times
    (h_0\omega^2+h_1\omega+h_2)
    \label{polynom}
\end{align}
with coefficients
\begin{align}
    q_0 &= (a-c a_1)(a-c a_2),
   \nonumber\\
   q_1 &= (a-c a_1)(b-d a_2)+(a-c a_2)(b-d a_1),
   \nonumber\\
   q_2 &= (b-d a_1)(b-d a_2),
   \nonumber\\
   h_0 &= (a-c a_3)(a-c a_4),
   \nonumber\\
   h_1 &= (a-c a_3)(b-d a_4)+(a-c a_4)(b-d a_3),
   \nonumber\\
   h_2 &= (b-d a_3)(b-d a_4).
    %\label{coeff}
    \nonumber
\end{align}
We choose  $a,b,c,d$ such that in polynomial  (\ref{polynom}) coefficients for $\omega^3$ and $\omega$ are nullified. Whence it follows conditions
\begin{eqnarray}
    q_0h_1+h_0q_1=0, \quad  q_1h_2+h_1q_2=0,
    %\label{coeff1}
    \nonumber
\end{eqnarray}
which hold for $q_1=h_1=0$.

We obtain therefore a system of equations:
\begin{align}
   &2-\left(\frac{d}{b}+\frac{c}{a}\right)(a_1+a_2)+2\frac{d}{b}\frac{c}{a}a_1a_2=0,
   \nonumber\\[4pt]
   &2-\left(\frac{d}{b}+\frac{c}{a}\right)(a_3+a_4)+2\frac{d}{b}\frac{c}{a}a_3a_4=0,
    \label{coeff2}
\end{align}
from which the expressions for $\frac{d}{b}+\frac{c}{a}$ and $\frac{d}{b} \frac{c}{a}$ can be obtained. From four variables  $a,b,c,d$ we can choose two as parameters and solve the system of equations for the remaining two.
Integral in Eq.~(\ref{Leg2}) can be written now as
\begin{equation}
   I=\int_{-\frac{b}{a}}^{-\frac{b-d}{a-c}}
   \frac{(a d - b c)\sigma(\omega)\mathrm{d}\omega}
   {\sqrt{(q_0\omega^2+q_2)(h_0\omega^2+h_2)}},
    \label{Leg3}
\end{equation}
and can be reduced to a combination of elliptic integrals of the first, second, and third orders  \cite{Hancock}.

We apply the above described approach to the integral in Eq.~(\ref{P_1/2_1})
\begin{equation}
    I= \int_{0}^{1} \frac{\mathrm{d}\eta}{\sqrt{p(\eta)}}\frac{p(\eta)}{(\eta-\eta_1)^2(\eta-\eta_2)^2}.
    \label{P_1/2_11}
\end{equation}
We set $R(\eta)=p(\eta)$, $S(\eta)=\frac{p(\eta)}{(\eta-\eta_1)^2(\eta-\eta_2)^2}$ and
\begin{equation}
       a_1=1+x_r, \quad a_2=1, \quad a_3=0, \quad a_4=-y_r.
    %\label{a14}
    \nonumber
\end{equation}
After applying the linear fractional transform (\ref{frac_lin}) with parameters $a=-1$ and $b=1$ (just a convenient choice),
we get solutions of the system (\ref{coeff2}):
\begin{align}
   c &= \frac{1+x_r+\sqrt{D}}{(1+x_r)y_r},\quad
   d=-\frac{1+x_r-\sqrt{D}}{(1+x_r)y_r},
   \nonumber\\
   c &= \frac{1+x_r-\sqrt{D}}{(1+x_r)y_r},\quad
   d =-\frac{1+x_r+\sqrt{D}}{(1+x_r)y_r},
    \label{coeff3}
\end{align}
where
\begin{equation}
      D=(1+x_r)(1+y_r)(1+x_r+y_r).
    \label{D}
\end{equation}
It is not important which pair to choose; we  take the second one form Eq.~(\ref{coeff3}), which guarantees  $ad-bc>0$.

Thus, we get
\begin{align}
     ad-bc &=\frac{2\sqrt{D}}{(1+x_r)y_r}, \quad -\frac{b}{a}=1,
     \nonumber\\
     -\frac{b-d}{a-c} & =\frac{(1+x_r)(1+y_r)+\sqrt{D}}{(1+x_r)(1+y_r)-\sqrt{D}},
    %\label{coeff4}
    \nonumber
\end{align}
and
\begin{align}
   q_{0,2} &=\frac{\left(1+x_r+y_r\mp\sqrt{D}\right)\!\left[(1+x_r)(1+y_r)\mp\sqrt{D}\right]}
   {(1+x_r)y_r^2},
   \nonumber\\
   h_{0,2} &=\pm\frac{\sqrt{D}}{1+x_r}.
    \label{q_h}
\end{align}
After substituting  (\ref{frac_lin}) and taking into account the above results, we arrive at
\begin{equation}
     p(\eta)=\frac{(q_0\omega^2+q_2)(h_0\omega^2+h_2)}{(c\omega+d)^4},
    %\label{p_omega}
    \nonumber
\end{equation}
\begin{align}
    q_0\omega^2+q_2 &= |q_0|\left(\frac{q_2}{|q_0|}-\omega^2\right),
    \nonumber\\
    h_0\omega^2+h_2 &= h_0(\omega^2-1),
    %\label{p_omega_add}
    \nonumber
\end{align}
\begin{align}
    \eta-\eta_1 &=\frac{2}{c\omega+d}\frac{(1+y_r)(1+x_r+y_r)+\sqrt{D}}{(2+x_r+y_r)y_r},
    \nonumber\\
    \eta-\eta_2 &=-\frac{2}{c\omega+d}\frac{(1+y_r)(1+x_r+y_r)-\sqrt{D}}{(2+x_r+y_r)y_r}\omega,
    %\label{eta_omega1}
    \nonumber
\end{align}

\begin{align}
     \sigma(\omega) &= \frac{(2+x_r+y_r)^2y_r^2}{16(1+y_r)^2(1+x_r+y_r)^2}|q_0|h_0
     \nonumber\\
     &\mathrel{\phantom{=}}
     \times \left(-\omega^2+\frac{q_2}{|q_0|}+1-\frac{q_2}{|q_0|\omega^2}\right).
    %\label
    \nonumber
\end{align}
Here we took into account that  $q_0<0$, $h_2<0$ and $h_2=-h_0$.
Reducing Eq.~\eqref{Leg3}, we get
\begin{align}
   I&=\theta\int_{1}^{\mu}
   \frac{\mathrm{d}\omega}
   {\sqrt{\left(\frac{q_2}{|q_0|}-\omega^2\right)\!\left(\omega^2-1\right)}}
   \nonumber\\
     &\mathrel{\phantom{=}}
     \times
   \left(-\omega^2\!+\!\frac{q_2}{|q_0|}+1-\frac{q_2}{|q_0|\omega^2} \right)\!,
    \label{Integ}
\end{align}
where
\begin{align}
    \theta&=\frac{(2+x_r+y_r)^2}{8D}
    \nonumber\\
     &\mathrel{\phantom{=}}
     \times\left[\sqrt{(1+x_r)(1+y_r)}-\sqrt{1+x_r+y_r}\right],
    \label{theta_ap}
\end{align}
\begin{equation}
     \mu=-\frac{b-d}{a-c}=\frac{(1+x_r)(1+y_r)+\sqrt{D}}{(1+x_r)(1+y_r)-\sqrt{D}}.
    \label{mu_ap}
\end{equation}

From Eq.~(\ref{q_h}) we obtain
\begin{equation}
      \frac{q_2}{|q_0|}=\frac{\left[\sqrt{D}+1+x_r+y_r\right]\!\left[(1+x_r)(1+y_r)+\sqrt{D}\right]}
      {\left[\sqrt{D}-(1+x_r+y_r)\right]\!\left[(1+x_r)(1+y_r)-\sqrt{D}\right]}.
    %\label{coeff5}
    \nonumber
\end{equation}
Next we use Eq.~(\ref{D})  and identity
\begin{equation}
 \frac{\lambda_1+\sqrt{\lambda_1\lambda_2}}{\lambda_1-\sqrt{\lambda_1\lambda_2}}=\frac{\sqrt{\lambda_1\lambda_2}+\lambda_2}{\sqrt{\lambda_1\lambda_2}-\lambda_2},
    %\label{tojd}
    \nonumber
\end{equation}
and find out
\begin{equation}
     \frac{(1+x_r)(1+y_r)+\sqrt{D}}{(1+x_r)(1+y_r)-\sqrt{D}}=\frac{\sqrt{D}+1+x_r+y_r}{\sqrt{D}-(1+x_r+y_r)}.
    %\label{tojd1}
    \nonumber
\end{equation}
Therefore, the following holds
\begin{equation}
     \mu=\frac{\sqrt{D}+1+x_r+y_r}{\sqrt{D}-(1+x_r+y_r)}, \quad \frac{q_2}{|q_0|}=\mu^2.
    \label{omega1}
\end{equation}

Integral~\eqref{Integ} takes a form
\begin{equation} 
   I=\theta\int_{1}^{\mu}
   \frac{\mathrm{d}\omega}
   {\sqrt{\left(\mu^2-\omega^2\right)\left(\omega^2-1\right)}}
   \left(-\omega^2+\mu^2+1-\frac{\mu^2}{\omega^2} \right).
   \nonumber
    %\label{Integ1}
\end{equation}
It is easy to see that $-\omega^2+\mu^2+1-\frac{\mu^2}{\omega^2}\geq0$ for $\omega\in [1,\mu]$, and therefore  $I>0$.

Finally, by using table integrals \cite{Prudnikov}, we get
\begin{equation}
   I=\theta\left[ \left(\mu+1/\mu\right)K\!\left(1\!-\!1/\mu^2\right)-2\mu E\!\left(1\!-\!1/\mu^2\right)\right]\!,
    \label{Integ_fin}
\end{equation}
where
\begin{align}
    K(m)&=\int_{0}^{\pi/2}\frac{\mathrm{d}\varphi}{\sqrt{1-m \sin^2\!\varphi}},
    \label{K_E1}
    %\nonumber
    \\
    E(m)&=\int_{0}^{\pi/2}\mathrm{d}\varphi \sqrt{1-m \sin^2\!\varphi}
    \label{K_E2}
\end{align}
are complete elliptic integrals of the first and second orders, respectively.
It is noteworthy that Eq.~\eqref{Integ_fin} can be recast by using the hypergeomtric functions, by using $K(m)=\frac{\pi}{2}{}_2F_1\left(\frac{1}{2},\frac{1}{2};1;m\right)$ and $E(m)=\frac{\pi}{2}{}_2F_1\left(-\frac{1}{2},\frac{1}{2};1;m\right)$.


\begin{thebibliography}{16}

\bibitem{yosi1} M.~F. Shlesinger,  J. Klafter, and  Y. Wong, \emph{Random walks with infinite spatial and temporal moments},  J. Stat. Phys.  \textbf{27}, 499 (1982).

\bibitem{yosi2} M.~F. Shlesinger, B.~J. West, and  J. Klafter, \emph{L\'{e}vy dynamics of enhanced diffusion: applications to turbulence}, Phys. Rev. Lett. \textbf{58}, 1100 (1987).

\bibitem{rmp}  V. Zaburdaev, S. Denisov, J. Klafter, \emph{L\'{e}vy walks}, Rev. Mod. Phys. \textbf{87}, 483 (2015).

\bibitem{prl2016} V. Zaburdaev, I. Fouxon, S. Denisov, and E. Barkai, \emph{Superdiffusive dispersals impart the geometry of underlying random walks}, Phys. Rev. Lett. \textbf{117}, 270601 (2016).

\bibitem{marcin2016} M. Magdziarz and
T. Zorawik, \emph{Explicit densities of multidimensional ballistic L\'{e}vy walks}, Phys. Rev. E \textbf{94}, 022130 (2016).

\bibitem{marcin2017} M. Magdziarz and
T. Zorawik, \emph{Method of calculating densities for isotropic ballistic L\'{e}vy walks}, Commun. Nonlinear Sci. Numer. Simul. \textbf{48}, 462 (2017).

\bibitem{fouxon2017} I. Fouxon, S. Denisov, V. Zaburdaev, and E. Barkai, \emph{Limit theorems for L\'evy walks in $d$ dimensions: rare and bulk fluctuations}, J. Phys. A: Math. Theor. \textbf{50}, 154002 (2017).

\bibitem{yosi3}  J. Klafter and G. Zumofen, \emph{L\'evy statistics in a Hamiltonian system}, Phys. Rev. E \textbf{49}, 4873 (1994).

\bibitem{dice}
L. van der Heijdt, \textit{Face to Face with Dice: 5000 Years of Dice and Dicing} (Gopher Publishers, Groningen, 2002).

\bibitem{zarfaty1} L. Zarfaty, A. Peletskyi, I. Fouxon, S. Denisov, and E. Barkai, \emph{Dispersion of particles in an infinite-horizon Lorentz gas}, Phys. Rev. E \textbf{98}, 010101 (2018).

\bibitem{cristadoro}  G. Cristadoro, T. Gilbert, M. Lenci, and D.~P. Sanders, \emph{Measuring logarithmic corrections to normal diffusion in infinite-horizon billiards}, Phys. Rev. E \textbf{90}, 022106 (2014).

\bibitem{zarfaty2} L. Zarfaty, A. Peletskyi, E. Barkai, and S. Denisov, \emph{Infinite horizon billiards: Transport at the border between Gauss and L\'evy universality classes}, Phys. Rev. E \textbf{100}, 042140 (2019).

\bibitem{bal1} D. Froemberg, M. Schmiedeberg, E. Barkai, and V. Zaburdaev, \emph{Asymptotic densities of ballistic L\'evy walks}, Phys. Rev. E \textbf{91}, 022131 (2015).

%\bibitem{ZDK2015}
%V. Zaburdaev, S. Denisov, and J. Klafter. L\'{e}vy walks // Rev. Mod. Phys. %\textbf{87}, 483 (2015).

%\bibitem{ZFDE2016}
%V. Zaburdaev, I. Fouxon, S. Denisov, and E. Barkai. Superdiffusive Dispersals Impart the Geometry of Underlying Random Walks // Phys. Rev. Lett. \textbf{117}, 270601 (2016).

%\bibitem{Feller2}
%W. Feller.  \emph{An introduction to probability theory and its applications. Vol. II} (John Wiley \& Sons Inc., NY, 1971).

%\bibitem{MBSB2002}
%M. M. Meerschaert, D. A. Benson, H.-P. Scheffler,  and B. Baeumer. Stochastic solution of space-time fractional diffusion equations // Phys. Rev. E \textbf{65}, 041103 (2002).

\bibitem{MBSB2002(r)}
M.~M. Meerschaert, D.~A. Benson, H.-P. Scheffler,  and P. Becker-Kern, \emph{Governing equations and solutions of anomalous random walk limits}, Phys. Rev. E \textbf{66}, 060102(R) (2002).

\bibitem{SM2003}
I.~M. Sokolov and R. Metzler, \emph{Towards deterministic equations for L\'{e}vy walks: The fractional material derivative}, Phys. Rev. E \textbf{67}, 010101 (2003).

\bibitem{MetzlerKlafter2004} 
R. Metzler and J. Klafter, \emph{The restaurant at the end of the random walk: Recent developments in the description of anomalous transport by fractional dynamics}, J. Phys. A: Math. Gen. \textbf{37}, R161 (2004).

\bibitem{GL2001}
C. Godr\`{e}che and J.~M. Luck, \emph{Statistics of the occupation time of renewal processes}, J. Stat. Phys. \textbf{104}, 489 (2001).

%\bibitem{GnedenkoKolmogorov}
%B.V. Gnedenko and A.N. Kolmogorov. Limit Distributions of Sums of Independent Random Variables // Cambridge, MA: Addison-Wesley (1954).

%\bibitem{Zolotarev}
%V.M. Zolotarev. One-dimensional stable distributions //
%Amer. Math. Soc., Vol. 65 of Transl. of Math. Monographs (1986).

\bibitem{SamorodnitskyTaqqu}
G. Samorodnitsky and M.~S. Taqqu, \emph{Random Processes: Stochastic Models with Infinite Variance Stable Non-Gaussian} (Chapman and Hall, NY, 1994).

\bibitem{Hancock}
H. Hancock, \emph{Lectures on the Theory of Elliptic Functions. Vol. I} (John Wiley \& Sons Inc., NY, 1910).

%\bibitem{Fedoryuk}
%%М. В. Федорюк. Асимптотика: интегралы и ряды // Москва: Наука. Гл. ред. физ-мат. лит. (1987).
%M.~V. Fedoryuk, \emph{Asymptotics: Integrals and Series} (Nauka, Moscow, 1987) [in Russian].

\bibitem{comment0} Formally, $x_r=x_t$ and $y_r=y_t$. However,  since variables  $x_r$ and $y_r$ are related to the expression which does not depend on $t$ explicitly, we use different notations here to distinguish these two different situations.


\bibitem{RalstonRabinowitz}
A. Ralston, P. Rabinowitz, \emph{A First Course in Numerical Analysis} (Dover Publ. Inc., NY, 2001).

\bibitem{comment1}
An alternative is to split  expression in Eq.~\eqref{Q12x} into summands of two different types, which do and do not include singular multiplier $(1-\eta)^{\gamma-1}$. However, in this case we would need two different types of orthogonal polynomials; e.g., for weights $w_j$ and nodes $\eta_j$.

\bibitem{to_be_published}
Yu.~S. Bystrik and S. Denisov, \emph{in preparation}.

\bibitem{Widder}
D.~V. Widder, \emph{The Laplace Transform}~(Princeton University Press, 1946).

\bibitem{Prudnikov}
A.~P. Prudnikov, Yu.~A. Brychkov, and O.~I. Marichev, \emph{ Integrals and Series, Volume 1: Elementary Functions} (Taylor \& Francis, London, 2002).

\bibitem{Vladimirov}
V.~S. Vladimirov, \emph{Equations of Mathematical Physics} (Dekker, NY, 1971).


\bibitem{asymptotics} Sections of the asymptotic pdf's along different directions can be evaluated analytically and, e.g., for $\gamma < 0.5$, two singularities  
going along the axes, power-law and logarithmic ones, can be distinguished~\cite{to_be_published}. 
%I. Fouxon, S. Denisov, V. Zaburdaev, and E. Barkai, J. Phys. A: Math. Theor. \bf{50} 154002 (2017).

\bibitem{covariant}
A. Rebenshtok, S. Denisov, P. H\"{a}nggi, and E. Barkai, \emph{Non-normalizable densities in strong anomalous diffusion: Beyond the central limit theorem}, Phys. Rev. Lett. \textbf{112}, 110601 (2014).

\end{thebibliography}
\end{document}